\documentclass[12pt]{article}
\usepackage{graphicx}
\usepackage{amsmath}

\newtheorem{theorem}{Theorem}[section]
\newtheorem{definition}[theorem]{definition}
\newtheorem{lemma}[theorem]{Lemma}

\newtheorem{corollary}[theorem]{Corollary}
\newtheorem{remark}[theorem]{Remark}
\newenvironment{proof}[1][Proof]{\textsc{#1.} }{\ \rule{0.5em}{0.5em}}
\numberwithin{equation}{section}
\input{tcilatex}

\begin{document}

\title{Einstein constraints on compact \\
n dimensional manifolds.}
\author{Yvonne Choquet-Bruhat}
\maketitle

\begin{abstract}
We give a general survey of the solution of the Einstein constraints by the
conformal method on n dimensional compact manifolds. We prove some new
results about solutions with low regularity (solutions in $H_{2}$ when n=3),
and solutions with unscaled sources.
\end{abstract}

\emph{Dedicated to Vincent Moncrief, a great scientist and friend.}

\bigskip

\section{Introduction.}

The geometric (physical) initial data on an n-dimensional manifold $M$ for
the Einstein equations are $\bar{g}$ a properly riemannian metric and $K$ a
symmetric 2-tensor. They cannot be arbitrary, they must satisfy the
constraints, which are the Gauss-Codazzi equations linking the metric $\bar{g%
}$ induced on $M$ by the space time metric $g,$ and the extrinsic curvature $%
K$ of $M$ as submanifold imbedded in the space time ($V,g$), with the value
on $M$ of the Ricci tensor of $g$.

These constraints read as equations on $M.$ They are the so called
hamiltonian constraint (we choose units such that 8$\pi G=1)$: 
\begin{equation}
R(\bar{g})-K.K+(trK)^{2}=2\rho ,
\end{equation}
and the momentum constraint: 
\begin{equation}
\bar{\nabla}.K-\bar{\nabla}trK=J.
\end{equation}
$\rho $ is a scalar and $J$ a vector on $M$ determined by the stress energy
tensor of the sources and the lapse $N$. In a Cauchy adapted frame, where
the equation of $M$ in $V$ is $x^{0}\equiv t=$constant, one has: 
\begin{equation*}
J_{i}=NT_{i}^{0},\text{ \ \ \ }\rho =N^{2}T^{00}
\end{equation*}
\smallskip with $\rho \geq 0$ if the sources satisfy a positive energy
condition.

In this article we use the standard method\footnote{{\footnotesize The
constraints were first formulated as an elliptic system in CB 1957,
explained also in CB 1962, through harmonic spacetime coordinates. The
decomposition of the constraints through the conformal method into a linear
system and an elliptic semi linear equation was obtained for zero trace K
and zero momentum sources by Lichnerowicz 1944. An elliptic formulation for
the linear system was given in CB 1970. The decomposition of the system of
constraints was generalized to constant trace K and scaled momentum by York
1972, who obtained an elliptic system by the now universally used conformal
splitting. Improved variants of this splitting, useful for numerical
computations, have been given by York 1999 and Pfeister and York 2002.}} to
solve the constraints, that is the conformal method, initiated by
Lichnerowicz and developped by CB and York. We treat the cosmological case%
\footnote{{\footnotesize The asymptotically euclidean case (isolated
systems) is considered in CB- Isenberg- York 2000. Results with similar low
regularity can be proved by analogous methods. They are being obtained
independently by D.\ Maxwell who also treats the case of asymptotically
euclidean manifolds with boundary..}}, that is we suppose the manifold $M$
to be compact (without boundary). We consider arbitrary dimensions and we
lower the regularity previously obtained for the solutions: we obtain large
classes of solutions in $H_{2}$ when $n=3.$ We prove also miscellaneous new
results. There is certainly still room for improvements.

\section{The conformal method.}

A riemannian metric $\gamma $ is arbitrarily chosen on the manifold $M,$
together with a scalar function $\tau .$ The physical metric $\bar{g}$ and
second fundamental form $K$ are then defined by\footnote{{\footnotesize In
the case n=2 the powers of }$\varphi ${\footnotesize \ \ are to be replaced
by exponentials (see Moncrief 1986)}$.$}: 
\begin{equation}
\bar{g}_{ij}\equiv \varphi ^{\frac{4}{n-2}}\gamma _{ij},\text{ \ \ }%
K^{ij}\equiv \varphi ^{\frac{-2n(n+2)}{n-2}}A^{ij}+\frac{1}{n}\bar{g}%
^{ij}\tau ,
\end{equation}
with $A^{ij}$ a traceless symmetric tensor. The sources are supposed to be
known, they split into York scaled and unscaled ones, they are given as a
pair of non negative scalars $\rho _{1},\rho _{2},$ and a pair of vectors $%
J_{1},J_{2}.$ The physical sources are: 
\begin{equation}
q\equiv \varphi ^{-\frac{2(n+1)}{n-2}}\rho _{1}+\rho _{2},\text{ \ \ }%
J\equiv \varphi ^{-\frac{2(n+1)}{n-2}}J_{1}+J_{2}.
\end{equation}
One denotes by $D,$ and by $\Delta _{\gamma },$ the covariant derivative,
and the Laplace operator, in the metric $\gamma .$

The Hamiltonian constraint becomes the Lichnerowicz equation: 
\begin{equation}
\mathcal{H}\equiv \Delta _{\gamma }\varphi -f(.,\varphi )=0,\text{ \ }
\end{equation}
\begin{equation*}
f(.,\varphi )\equiv r\varphi -a\varphi ^{-\frac{3n-2}{n-2}}-q_{1}\varphi ^{-%
\frac{n}{n-2}}+(b-q_{2})\varphi ^{\frac{n+2}{n-2}},
\end{equation*}
with (a dot is the scalar product in $\gamma )$%
\begin{equation*}
\text{\ }r\equiv \frac{n-2}{4(n-1)}R(\gamma )\text{\ \ \ }a\equiv \frac{n-2}{%
4(n-1)}A.A\geq 0,\text{ \ }b\equiv \frac{n-2}{4n}\tau ^{2}\geq 0,
\end{equation*}
\begin{equation*}
\text{ \ }q_{i}\equiv \frac{n-2}{2(n-1)}\rho _{i}\geq 0.
\end{equation*}

The momentum constraint becomes (indices raised with $\gamma )$: 
\begin{equation}
\mathcal{M}^{i}\equiv D_{j}A^{ij}-\{\frac{n-1}{n}\varphi ^{\frac{2n}{n-2}%
}\partial ^{i}\tau +\varphi ^{\frac{2(n+2)}{n-2}}J_{2}+J_{1}\}=0,
\end{equation}
In the original conformal formulation one sets: 
\begin{equation}
A^{ij}\equiv B^{ij}+(\mathcal{L}_{\gamma ,conf}X)^{ij}
\end{equation}
with $B^{ij}$ an arbitrary given traceless tensor, and $\mathcal{L}_{\gamma
,conf}X$ the conformal Lie derivative of a vector $X$ to be determined, that
is: 
\begin{equation}
(\mathcal{L}_{\gamma ,conf}X)_{ij}\equiv D_{i}X_{j}+D_{j}X_{i}-\frac{2}{n}%
\gamma _{ij}D_{h}X^{h}.
\end{equation}
We will call the system 2.3 and 2.4, like Isenberg and Moncrief, the LCBY
equations. The unknowns are $\varphi $ and $X.$

\begin{remark}
In the thin sandwich formulation (York 1999), which makes the conformal
invariance of the constraints (when $\tau =$constant) more transparent and
is useful in numerical computations, one sets: 
\begin{equation}
A^{ij}\equiv (2N)^{-1}\{-u^{ij}+(\mathcal{L}_{\gamma ,conf}\beta )^{ij}\}
\end{equation}
with $N$ an arbitrary, positive, scalar, and $u^{ij}$ an arbitrary given
traceless tensor. The momentum contraint in this formulation reads, for
general $n$: 
\begin{equation}
\mathcal{M}^{i}\equiv D_{j}\{(2N)^{-1}(\mathcal{L}_{\gamma ,conf}\beta
)^{ij}\}-F_{TS}^{i}(.,\varphi )=0,
\end{equation}
\begin{equation*}
F_{TS}^{i}(.,\varphi )\equiv D_{j}\{(2N)^{-1}u^{ij}\}+\frac{n-1}{n}\varphi
^{2n/n-2}\partial ^{i}\tau +\varphi ^{2(n+2)/(n-2)}J_{2}+J_{1}.
\end{equation*}
The unknowns are now $\beta $ and $\varphi .$ The mathematical results are
essentially the same as in the original formulation which we use here for
simplicity of notations.
\end{remark}

\textbf{Functional spaces.}

To have good functional spaces it is convenient to introduce a given smooth
properly riemannian metric $e$ on $M.$

The Sobolev spaces $W_{s}^{p}$ on $(M,e)$ are defined as closures of spaces
of smooth tensor fields in the norm 
\begin{equation}
||f||_{W_{s}^{p}}\equiv \{\int_{M}\sum_{0\leq k\leq s}|\partial
^{k}f|^{p}\mu _{e}\}^{\frac{1}{p}},
\end{equation}
where $\partial ,$ $|.|$ and $\mu _{e}$\ are the covariant derivative, the
pointwise norm and the volume element in the metric $e.$ A metric $\gamma $
is said to belong to $M_{\sigma }^{p}$ if it is properly riemannian and $%
\gamma \in W_{\sigma }^{p}.$ We will always suppose that $\sigma >\frac{n}{p}%
,$ then $\gamma \in M_{\sigma }^{p}$ is continuous on $M$ as well as its
contrevariant associate $\gamma ^{\#},$ and $M_{\sigma }^{p}$ is an open set
in $W_{\sigma }^{p}.$ The volume element $\mu _{\gamma }$ of $\gamma $ is
equivalent to $\mu _{e}.$

\section{Solution of the momentum constraint.}

The momentum constraint 2.4 reads: 
\begin{equation}
(\Delta _{\gamma ,conf}X)^{i}\equiv D_{j}(\mathcal{L}_{\gamma
,conf}X)^{ij}=F^{i},
\end{equation}
\begin{equation}
F^{i}\equiv \frac{n-1}{n}\varphi ^{2n/(n-2)}\partial ^{i}\tau +\varphi
^{2(n+2)/(n-2)}J_{2}+J_{1}-D_{i}B^{ij}
\end{equation}
where $B$ is a traceless symmetric 2-tensor. When $\gamma ,$ properly
riemannian, $\tau ,$ $J_{1},$ $J_{2},$ and $B$ are given and $\varphi $ is
known $3.1$ is a linear system for the vector $X.$

\begin{lemma}
If $\gamma \in M_{2+s}^{p},$ $s\geq 0,$ $p>\frac{n}{2}$ the vector $F$
belongs to $W_{s}^{p},$ as soon as $\varphi \in W_{2+s}^{p},$ $D\tau ,J\in
W_{s}^{p}$ and $B\in W_{s+1}^{p}.$
\end{lemma}

\begin{proof}
The Sobolev multiplication theorem.
\end{proof}

\begin{theorem}
The momentum constraint 3.1 with $\gamma \in M_{2+s}^{p},$ $p>\frac{n}{2},$ $%
s\geq 0$ has a solution $X\in W_{2+s}^{p},$ if $F\in W_{s}^{p}$ and is $%
L^{2} $ orthogonal to the space of conformal Killing (CK) vector fields of ($%
M,\gamma ).$

The solution is determined up to addition of a CK vector. It is unique if we
impose that it be $L^{2}$ orthogonal to CK vectors. There exists then a
constant $C_{\gamma }>0$ depending only on $\gamma $ such that 
\begin{equation}
\parallel X\parallel _{W_{s+2}^{p}}\leq C_{\gamma }\parallel F\parallel
_{W_{s}^{p}}.
\end{equation}
\end{theorem}

\begin{proof}
We show that the operator $\Delta _{\gamma ,conf}$ satisfies the theorems
7.2 and 7.4 of the appendix A.

1. The operator is elliptic: its principal symbol at $x$, with $\xi \in
T_{x}M$ is the linear mapping from covariant vectors $b$ into covariant
vectors $a$ given by 
\begin{equation}
\xi ^{i}\xi _{i}b_{j}+\xi ^{i}\xi _{j}b_{i}-\frac{2}{n}\xi _{j}\xi
^{k}b_{k}=a_{j}\;.
\end{equation}
This linear mapping is an isomorphism if $\xi \neq 0:$ its characteristic
determinant is computed to be ($\gamma ^{\#}$ is the contrevariant tensor
associated with $\gamma ):$%
\begin{equation}
D(\xi )\equiv (\xi ^{i}\xi _{i})^{n}(1-\frac{1}{n})\equiv \gamma ^{\#}(\xi
,\xi )>0\text{ \ if \ }\xi \not=0.
\end{equation}

2.\ The coefficients of the operator $\Delta _{\gamma ,conf}$ are of the
type $:$%
\begin{equation}
a_{2}\tilde{=}\gamma ^{\#},\text{ \ }a_{1}\tilde{=}\gamma ^{\#}\partial
\gamma ,\text{ \ }a_{0}\tilde{=}\gamma ^{\#}\partial \gamma \partial \gamma
+\gamma ^{\#}\partial ^{2}\gamma
\end{equation}
they satisfy the hypothesis of the theorem 7.2 of the appendix A if 
\begin{equation}
a_{2}\in M_{2}^{p},\text{ \ }a_{1}\in W_{_{1}}^{p},\text{ \ }a_{0}\in L^{p}%
\text{ \ with \ }p>\frac{n}{2}\text{\ .}
\end{equation}
Indeed if $\gamma \in M_{2}^{p},p>\frac{n}{2},$ then also $\gamma ^{\#}\in
M_{2}^{p}\subset C^{0,\alpha },$ while $\ \partial \gamma \partial \gamma
\in W_{1}^{p}\times W_{1}^{p}\subset L^{p},$ hence $a_{1}\in W_{1}^{p},$ and 
$a_{0}\in L^{p}.$

2. Smooth vectorfields $X$ and $Y$ with smooth $\gamma $ satisfy, on a
compact manifold, the following identity: 
\begin{equation}
\int_{M}Y_{j}(\Delta _{\gamma ,conf}X)^{j}\mu _{\gamma }\equiv
\int_{M}Y_{j}D_{i}\left( D^{i}X^{j}+D^{j}X^{i}-\frac{2}{n}\gamma
^{ij}D_{k}X^{k}\right) \mu _{\gamma }\equiv
\end{equation}
\begin{equation*}
-\int_{M}\left( D^{i}Y^{j}+D^{j}Y^{i}-\frac{2}{n}\gamma
^{ij}D_{k}Y^{k}\right) \left( D_{i}X_{j}+D_{j}X_{i}-\frac{2}{n}\gamma
_{ij}D_{l}X^{l}\right) \mu _{\gamma }
\end{equation*}
Under the hypothesis made on $p$ the operator $\Delta _{\gamma ,conf}$ is a
continuous mapping $W_{2}^{p}\rightarrow L^{p},$ hence $Y.\Delta _{\gamma
,conf}X$ is a continuous mapping $W_{2}^{p}\times W_{2}^{p}\rightarrow
L^{1}. $ The conformal Killing operator is of the form $\partial +\gamma
^{\#}\partial \gamma ,$ it is a continuous mapping from $W_{2}^{p}$ into $%
W_{1}^{p},$ and by the Sobolev embedding it holds that $W_{1}^{p}\times
W_{1}^{p}\subset L^{1}$ as soon as $p\geq \frac{2n}{n+2},$ a fortiori if $p>%
\frac{n}{2}.$

The continuity of all the considered embeddings permits the passage to the
limit which proves the identity 3.8 in our low regularity case. This
identity implies, making $X=Y,$ $\Delta _{\gamma ,conf}X=0,$ the following
one: 
\begin{equation}
(\mathcal{L}_{\gamma ,conf}X)_{ij}\equiv D_{i}X_{j}+D_{j}X_{i}-\frac{2}{n}%
\gamma _{ij}D_{l}X^{l}=0.
\end{equation}

The identity 3.8 also shows that $\Delta _{\gamma
,conf}-k:W_{2}^{p}\rightarrow L^{p}$ is injective, hence an isomorphism, if $%
k$ is a strictly positive number. Its inverse is then a compact operator,
hence $\Delta _{\gamma ,conf}$ is a Fredholm operator, $\Delta _{\gamma
,conf}X=F$ has a solution $X\in W_{2}^{p}$ iff $F$ is orthogonal to the
kernel of the adjoint operator, which is $\Delta _{\gamma ,conf}$ itself.
The solution is unique in the Banach space of $W_{2}^{p}$ vectors orthogonal
to CK vectors.

The proof of higher regularity for more regular coefficients, and of the
inequality 3.3\ is obtained by derivating the equation.
\end{proof}

\begin{remark}
If the Ricci tensor\footnote{{\footnotesize The hypothesis }$\gamma \in
M_{2}^{p},$ {\footnotesize p\TEXTsymbol{>}n/2\ \ implies that the Riemann
and Ricci tensors of }$\gamma ${\footnotesize \ are in L}$^{p}.$} of $\gamma 
$, with components $\rho _{ij},$ is negative definite the manifold $%
(M,\gamma )$ admits no conformal Killing fields. Indeed the equality $\Delta
_{\gamma ,conf}X=0$ implies, using the Ricci identity, 
\begin{equation}
D^{i}D_{i}X_{j}+(1-\frac{2}{n})D_{j}D^{i}X_{i}+\rho _{jl}X^{l}=0,
\end{equation}
which implies on a compact manifold 
\begin{equation}
\int_{M}\{-D^{i}X^{j}D_{i}X_{j}-(1-\frac{2}{n})D_{j}X^{j}D^{i}X_{i}+\rho
_{jl}X^{l}X^{j}\}\mu _{\gamma }=0,
\end{equation}
from which follows $X\equiv 0$ if Ricci($\gamma )$ is negative definite and $%
n\geq 2.$
\end{remark}

\begin{remark}
Even when $X$ is not unique (i.e. if $(M,\gamma )$ admits a C.K vector
field) the tensor $A$ is determined uniquely through the formula 2.5.
\end{remark}

\begin{lemma}
When $D\tau \equiv 0$ and $J_{2}\equiv 0,$ then $F$ is $L^{2}$ orthogonal to
CK vector fields if and only if it is so of $J_{1}.$
\end{lemma}

\begin{proof}
If $D\tau \equiv 0$ and $J_{2}\equiv 0,$ then $F$ reduces to the sum of $%
J_{1}$ and the divergence of a traceless symmetric tensor. Such divergences
are $L^{2}$ orthogonal to CK vector fields. This fact well known in a
smoother case results from the identity, still valid under our hypotheses 
\begin{equation}
\int_{M}D_{i}B^{ij}X_{j}\mu _{\gamma }=-\int_{M}B^{ij}(\mathcal{L}_{\gamma
,conf}X)_{ij}\mu _{\gamma }.
\end{equation}
\end{proof}

If $D\tau \not\equiv 0$ or $J_{2}\not\equiv 0$ (unscaled momentum sources)
there seems to be at present no result for manifolds $(M,\gamma )$ admitting
CK vector fields.

\section{Lichnerowicz equation.}

It reads 
\begin{equation}
\Delta _{\gamma }\varphi =f(.,\varphi )\equiv r\varphi -a\varphi ^{-\frac{%
3n-2}{n-2}}-q_{1}\varphi ^{-\frac{n}{n-2}}+(b-q_{2})\varphi ^{\frac{n+2}{n-2}%
},
\end{equation}
we consider in this section that the coefficients $a,q_{i},b$ are given non
negative functions on $M.$

\subsection{The Yamabe properties.}

A theorem conjectured by Yamabe and proved by Trudinger, Aubin, Schoen is
interesting to classify solutions obtained by the conformal method.

\begin{definition}
The functional 
\begin{equation}
J_{\gamma }(\varphi )\equiv \frac{\int_{M}(k_{n}^{-1}|\nabla \varphi
|^{2}+R(\gamma )\varphi ^{2})\mu _{\gamma }}{(\int_{M}\varphi ^{2n/(n-2)}\mu
_{\gamma })^{(n-2)/n}}.
\end{equation}
defined for every $\varphi \in H_{1}\subset L^{\frac{2n}{n-2}},\ \varphi
\not\equiv 0$ and $\gamma \in M_{2}^{p}$ $p>\frac{n}{2},$ is called the
Yamabe functional.
\end{definition}

The functional $J_{q}(\varphi )$ admits an infimum for $\varphi \in
W_{2}^{p} $ (dense in $H_{1}$ if $p>\frac{n}{2}),$ and $\varphi \not\equiv
0, $ because it is bounded below as shown by the inequality: 
\begin{equation}
\int_{M}(k_{n}^{-1}|\nabla \varphi |^{2}+R(\gamma )\varphi ^{2})\mu _{\gamma
}\geq -|\int_{M}R(\gamma )\varphi ^{2}\mu _{\gamma }|\geq -||R(\gamma
)||_{L^{\frac{n}{2}}}||\varphi ^{2}||_{L^{\frac{n}{n-2}}}
\end{equation}
This infimum $\mu $: 
\begin{equation}
\mu \equiv \underset{\varphi \in W_{2}^{p},\varphi \not\equiv 0}{Inf}%
J_{\gamma }(\varphi )\equiv \underset{\varphi \in W_{2}^{p},\varphi \geq
0,\varphi \not\equiv 0}{Inf}J_{\gamma }(\varphi ),
\end{equation}
depends only on the conformal class of $\gamma ,$ it is called the Yamabe
invariant. The manifolds $(M,\gamma )$ are split into 3 \textbf{Yamabe
classes }according to the sign of $\mu .$

\begin{definition}
The manifold $(M,\gamma )$ is said to be in the negative Yamabe class if $%
\mu <0,$ in the zero Yamabe class if $\mu =0$, in the positive Yamabe class
if $\mu >0.$
\end{definition}

\begin{remark}
It is known (Kazdan and Warner (1975, 1985) that a sufficient condition for $%
(M,\gamma )$ to be of negative Yamabe class is $\int_{M}R(\gamma )\mu
_{\gamma }<0,$ and that every compact manifold of dimension $n\geq 3$ admits
a metric in the negative Yamabe class. On the other hand, not all manifolds
(Lichnerowicz 1963) can support metrics in the positive or zero Yamabe class.
\end{remark}

Possible topologies of manifolds of various Yamabe type are reviewed in
Fisher and Moncrief 1994.

The following theorem conjectured by Yamabe has been proved in an increasing
number of cases by Trudinger and Aubin and finally completed by Schoen, by
showing that the infimum is attained by a function $\varphi _{m}.$

\begin{theorem}
Let $M$ be a compact smooth manifold. Any smooth riemannian metric $\gamma $
on $M$ is conformal to a metric with constant scalar curvature, the sign of
which is a conformal invariant.
\end{theorem}

\begin{remark}
The Yamabe theorem says that the minimum $\varphi _{m},$ hence $\gamma _{m},$
is unique if $\mu <0,$ but not necessarily so if $\mu >0.$
\end{remark}

The Yamabe invariant is defined as soon as $\gamma \in M_{2}^{p},$ $p>\frac{n%
}{2},$ since then $R(\gamma )\in L^{p},$ and $\varphi \in W_{2}^{p}\subset
C^{0}.$ Checking of the different steps of the proof shows that the
existence of $\varphi _{m}\in W_{2}^{p},$ $\varphi _{m}>0$ can be extended
with no real difficulty to these low regularity metrics in the cases $\mu <0$
and $\mu =0.$ Difficulties for this generalization appear in the case $\mu
>0 $ and specially when $n=3.$ The proof given by Schoen uses indeed the
geodesic balls and the Green function which require at least a $C^{1,1}$
metric, implied only by $\gamma \in W_{2+s}^{p}$, $p>\frac{n}{2},$ if $s\geq
2.$ The weaker theorem proved by Yamabe\footnote{{\footnotesize See for
instance Aubin p.127.}} extends to $\gamma \in M_{2}^{p},$ $p>\frac{n}{2}$
and will be sufficient for our use.

\begin{theorem}
Let $(M,\gamma )$ be a compact riemannian manifold with $\gamma \in
M_{2}^{p},$ $p$\TEXTsymbol{>}$\frac{n}{2}.$ Then:

1. If $\gamma $ is in the negative Yamabe class it is conformal to a metric
with scalar curvature $-1$.

2. If $\gamma $ is in the zero Yamabe class it is conformal to a metric with
scalar curvature 0.

3. If $\gamma $ is in the positive Yamabe class if it is conformal to a
metric with continuous and positive scalar curvature $r\geq 1.$
\end{theorem}

\subsection{Non existence and uniqueness.}

\begin{definition}
Let $f\geq 0$ be a function defined almost everywhere on $M.$ We say that%
\footnote{{\footnotesize If }$a${\footnotesize \ is continuous the
definition coincides with the usual notation.}} $f\not\equiv 0$ if there is
an open set of $M$ where $Inf$ $f>0.$
\end{definition}

\begin{theorem}
\textbf{\ }(non existence) The Lichnerowicz equation admits no solution $%
\varphi >0,$ $\varphi \in $ $W_{2}^{p},$ $p>\frac{n}{2},$ on a compact
manifold (M,$\gamma ),$ $\gamma \in M_{2}^{p},$ $a,b,q\in L^{1}$ if either:

1. $r\leq 0$, $b-q_{2}\leq 0,$ and $-r+a+q_{1}-b+q_{2}\not\equiv 0.$

2. $r\geq 0$, $a\equiv q_{1}\equiv 0,b-q_{2}\geq 0$ and $r+b-q_{2}\not\equiv
0.$
\end{theorem}

\begin{proof}
The integral of $\Delta _{\gamma _{n}}\varphi _{n}$ with respect to the
volume element of $\gamma _{n}$ is equal to zero on a compact manifold $M$
if $\gamma _{n}$ is $C^{1}$ and $\varphi _{n}$ is $C^{2},$ by the Stokes
formula. We approximate the given $\gamma $ and $\varphi $ in $%
W_{2}^{p}\subset C^{0}$ by such $\gamma _{n}$ and $\varphi _{n}.$ Denote by $%
\mu _{\gamma _{n}}$ and $\mu _{\gamma }$ the volume elements of respectively 
$\gamma _{n}$ and $\gamma ,$ $\mu _{\gamma _{n}}$ tends in $C^{0}$ to $\mu
_{\gamma }.$ The Sobolev embedding and multiplication theorems show that $%
\Delta _{\gamma }\varphi (\mu _{e})^{-1}\mu _{\gamma }-\Delta _{\gamma
_{n}}\varphi (\mu _{e})^{-1}\mu _{\gamma _{n}}$ tends to zero in $L^{1}$
(see analogous proofs in the appendix), hence the integral of $\Delta
_{\gamma }\varphi $ on $(M,\gamma )$ is zero.

On the other hand, $f(x,\varphi )$ is integrable on $M$ since $\varphi \in
C^{0}$ and $\varphi >0$ and its integral cannot vanish under the condition 1
[respectively 2] which implies that $f(x,\varphi )\leq 0$ [respectively $%
\geq 0]$ and is strictly negative [positive] on an open set of $M.$
\end{proof}

The geometrical origin of the Lichnerowicz equation leads to a general
uniqueness theorem, independant of the sign of $r,$ hence of the Yamabe
class of $\gamma $.

\begin{theorem}
(uniqueness\footnote{{\footnotesize First proved (case n=3, smooth
coefficients) by Araki, see CB 1962.}}) The Lichnerowicz equation 4.1 on $%
(M,\gamma ),$ $\gamma \in M_{2}^{p},p>\frac{n}{2},$ has at most one positive
solution $\varphi \in W_{2}^{p},$ if $a,q_{i},b\in L^{1},$ $a\geq 0$, $%
q_{1}\geq 0,$ $b-q_{2}\geq 0$, and $a+$ $q_{1}+$ $b-q_{2}\not\equiv 0$ on M$%
. $
\end{theorem}

\begin{proof}
Suppose it admits two solutions $\varphi _{1}>0$ and $\varphi _{2}>0$. The
following identity holds: 
\begin{equation}
\Delta _{\varphi _{2}^{4/(n-2)}\gamma }(\varphi _{1}\varphi
_{2}^{-1})-r(\varphi _{2}^{4/(n-2)}\gamma )(\varphi _{1}\varphi
_{2}^{-1})\equiv -(\varphi _{1}\varphi _{2}^{-1})^{(n+2)/(n-2)}r(\varphi
_{1}^{4/(n-2)}\gamma )
\end{equation}
\smallskip Since $\varphi _{1}$ is a solution of 4.1 we have an equality of
the form: 
\begin{eqnarray*}
r(\varphi _{1}^{4/(n-2)}\gamma ) &\equiv &-\varphi
_{1}^{-(n+2)/(n-2)}\{\Delta _{\gamma }\varphi _{1}-\varphi _{1}r(\gamma )\}
\\
&=&\varphi _{1}^{-(n+2)/(n-2)}\{a\varphi _{1}^{P}+q_{1}\varphi
^{P_{1}}-\varphi ^{Q}(b-q_{2})\}
\end{eqnarray*}
\smallskip and an analogous equation for $r(\varphi _{2}^{4/(n-2)}\gamma ).$
Inserting in the previous equation gives the equation: 
\begin{equation}
\Delta _{\varphi _{2}^{2q}\gamma }(\varphi _{1}\varphi _{2}^{-1}-1)-\lambda
\{(\varphi _{1}\varphi _{2}^{-1}-1\}=0
\end{equation}
with 
\begin{eqnarray*}
\lambda &\equiv &a\varphi _{1}^{(-3n+2)/(n-2)}\varphi _{2}^{-(n+2)/(n-2)}{%
\frac{(\varphi _{1}\varphi _{2}^{-1})^{4(n-1)/(n-2)}-1}{\varphi _{1}\varphi
_{2}^{-1}-1}}+ \\
&&q_{1}\varphi _{1}^{-n/(n-2)}\varphi _{2}^{-(n+2)/(n-2)}+(b-q_{2})\varphi
_{1}\varphi _{2}^{-1}{\frac{(\varphi _{1}\varphi _{2}^{-1})^{4/(n-2)}-1}{%
\varphi _{1}\varphi _{2}^{-1}-1}}
\end{eqnarray*}
The fractions with denominator $\varphi _{1}\varphi _{2}^{-1}-1$ are
continuous and positive functions on $M$ if it is so of the $\varphi
^{\prime }s,$ because of the mean function theorem which says that if $a$
and $\alpha $ are two positive numbers there exist a number $b$ such that 
\begin{equation}
\frac{a^{\alpha }-1}{a-1}=\alpha b^{\alpha -1},\text{ \ }b\in \lbrack a,1]%
\text{ \ if \ }0<a\leq 1,\text{ \ \ \ }b\in \lbrack 1,a]\text{ \ if \ }a\geq
1.
\end{equation}
Therefore $\lambda $ is an integrable function on $M$, depending on $\varphi
_{1}$, $\varphi _{2}$, $a,q_{1},q_{2}$ and $b$, with $\lambda \geq 0$ and $%
\lambda \not\equiv 0$ on $M$ if $a+q_{1}+b-q_{2}\geq 0$ and $\not\equiv 0.$
The equation implies therefore\footnote{{\footnotesize See appendix A, lemma
7.6.}} $\varphi _{1}\varphi _{2}^{-1}-1\equiv 0.$
\end{proof}

\subsection{\noindent Existence theorem, scaled sources in $L^{\infty }$.}

The results can be obtained, as in the case $n=3,$ by using the Leray
Schauder degree, CB 1972 (unscaled sources), O'Murchada and York 1974
(scaled sources), or a constructive method, Isenberg 1987. Both methods use
sub and supersolutions. The following theorem is an extension of results
obtained before with higher regularity, and for $n=3$ (see reviews in
CB-York 1980, Isenberg 1995). In the theorems some coefficients are supposed
to be in $L^{\infty },$ and not just in $L^{p},$ because we look for
constant sub and supersolutions. A new result is an extension to cases where 
$\tau ^{2}\not\equiv 0$ in the cases of negative or zero Yamabe class.

\begin{theorem}
The Lichnerowicz equation with scaled sources on a compact n-manifold (M,$e)$
with given riemannian metric $\gamma \in M_{2}^{p},$ $p>\frac{n}{2},$ and $a 
$, $b,$ $q_{1}\in L^{\infty },$ with unknown the conformal factor $\varphi ,$
admits a solution $\varphi >0,$ $\varphi \in W_{2}^{p}$ if:

1. ($M,\gamma )$ is in the positive Yamabe class and $\underset{M}{Inf}%
(a+q_{1})>0$.

2. ($M,\gamma )$ is in the zero Yamabe class, $\underset{M}{Inf}\tau ^{2}>0$
and $\underset{M}{Inf}(a+q_{1})>0$.

3.($M,\gamma )$ is in the \ negative Yamabe class and $\underset{M}{Inf}\tau
^{2}>0.$
\end{theorem}

\begin{proof}
The general existence theorem 8.2 shows that, under the hypothesis made on
the coefficients, the Lichnerowicz equation admits a solution $\varphi >0,$ $%
\varphi \in W_{2}^{p},$ if we can find constants $\ell $ and $m$ such that
on $M$: 
\begin{equation}
f(.,\ell )\leq 0,\text{ \ }f(.,m)\geq 0\text{ \ \ \ with \ \ }0<\ell \leq m.
\end{equation}
\smallskip We have: 
\begin{equation}
f(x,y)\equiv y^{-\frac{3n-2}{n-2}}h(x,y),\ 
\end{equation}
\begin{equation}
h(x,y)\equiv b(x)y^{\frac{4n}{n-2}}+r(x)y^{\frac{4(n-1)}{n-2}}-q_{1}(x)y^{%
\frac{2(n-1)}{n-2}}-a(x)
\end{equation}
There exist $\ell >0$ and $m\geq \ell $ satisfying 4.8 if they satisfy the
same inequalities with $h.$ We have 
\begin{equation}
h_{y}^{\prime }\equiv \frac{2}{n-2}\tilde{h}(x,y)y^{\frac{n}{n-2}},\text{ \ }%
\tilde{h}(x,y)\equiv 2nby^{^{\frac{2n+2}{n-2}}}+2(n-1)ry^{\frac{2n-2}{n-2}%
}-(n-1)q_{1}.
\end{equation}
\begin{equation}
\tilde{h}_{y}^{\prime }\equiv \frac{8}{n-2}\{n(n+1)by^{^{\frac{4}{n-2}%
}}+(n-1)^{2}r\}y^{\frac{n}{n-2}}.
\end{equation}
The function $\tilde{h}_{y}^{\prime }(x,.)$ has at most one zero $%
Y_{0}(x)>0; $ the function $\tilde{h}(x,.)$ starts from $-(n-1)q_{1}(x)\leq
0 $ for $y=0.$ We examine the various Yamabe classes.

\begin{itemize}
\item  1. $(M,\gamma )$ is of positive Yamabe class, we can suppose $\ r$
continuous and $r(x)\geq 1$ at every point $x\in M.$ Then for any given $%
x\in M$ the function $h_{y}^{\prime }(x,.)$ is a non decreasing function of $%
y\geq 0,$ non negative as soon as 
\begin{equation}
y\geq m(x),\text{ \ with \ }m(x)\text{ a number\ such that \ \ }m(x)^{\frac{%
2n-2}{n-2}}\geq \frac{1}{2}q_{1}(x).
\end{equation}
The function $h(x,.)$ of $y$ is non decreasing when $y\geq m(x),$ it is non
negative if it is so for $y=m(x).$ A sufficient condition is that, since $%
b\geq 0,$: 
\begin{equation}
m(x)^{\frac{4(n-1)}{n-2}}\geq q_{1}(x)m(x)^{\frac{2(n-1)}{n-2}}+a(x)
\end{equation}
which is implied for instance by the sufficient condition 
\begin{equation}
m(x)\geq Max\{1,a(x)+q_{1}(x)\}
\end{equation}
\end{itemize}

Any number $m$ (independent of $x$) such that 
\begin{equation}
m\geq \underset{M}{Sup}(a+q_{1}),\text{ \ \ }m\geq 1
\end{equation}
will satisfy the condition 4.15 almost everywhere on $M,$ hence will be a
supersolution.

The function $h(x,.)$ is non positive for some $\ell (x)$ if 
\begin{equation}
b(x)\ell (x)^{\frac{4n}{n-2}}+r(x)\ell (x)^{\frac{4(n-1)}{n-2}}-q_{1}(x)\ell
(x)^{\frac{2(n-1)}{n-2}}-a(x)\leq 0
\end{equation}
A sufficient condition is 
\begin{equation}
\ell (x)\leq 1,\text{ }\ \ \text{(}b(x)+r(x))\ell ^{\frac{4(n-1)}{n-2}}\leq
(q_{1}(x)+a(x))\ell ^{\frac{2(n-1)}{n-2}}
\end{equation}
a sufficient condition is therefore: 
\begin{equation}
\ell (x)\leq 1,\text{ }\ \ \ell (x)\leq \text{(}%
b(x)+r(x))^{-1}(q_{1}(x)+a(x))
\end{equation}
Therefore if $\underset{M}{Inf}(a+q_{1})>0$ any number $\ell $ such that 
\begin{equation}
0<\ell \leq \{\underset{M}{Sup}\text{(}b+r)\}^{-1}\{\underset{M}{Inf}%
(q_{1}+a)\},\text{ \ }\ell \leq 1.
\end{equation}
is a constant subsolution,and it is possible to choose $\ell $ and $m$ such
that 
\begin{equation}
0<\ell \leq m.
\end{equation}

\begin{itemize}
\item  2. $(M,\gamma $) is of zero Yamabe class, we suppose $r(\gamma )=0$.
Then: 
\begin{equation}
h(x,y)\equiv b(x)y^{\frac{4n}{n-2}}-q_{1}(x)y^{\frac{2(n-1)}{n-2}}-a(x)
\end{equation}
There exist $\ell >0$ and $m\geq \ell $ satisfying 4.8 if they satisfy the
same inequalities with $h.$ We have 
\begin{equation}
h_{y}^{\prime }\equiv \frac{2}{n-2}\tilde{h}(x,y)y^{\frac{n}{n-2}},\text{ \ }%
\tilde{h}(x,y)\equiv 2nby^{^{\frac{2n+2}{n-2}}}-(n-1)q_{1}.
\end{equation}
\begin{equation}
\tilde{h}_{y}^{\prime }\equiv \frac{4}{n-2}n(n+1)by^{^{\frac{n+4}{n-2}}}.
\end{equation}
\end{itemize}

There exist constant supersolutions $0<\ell \leq m$ if $\underset{M}{Inf}%
(a+q_{1},b)>0$ and $\underset{M}{Sup}(a+q_{1},b)<+\infty .$ They are then
chosen such that: 
\begin{equation}
0<\ell \leq Min\{1,\frac{\underset{M}{Inf}(a+q_{1})}{\underset{M}{Sup}b}\},%
\text{ \ \ \ }m\geq Max\{1,\frac{\underset{M}{Sup}(a+q_{1})}{\underset{M}{Inf%
}b}.
\end{equation}

\begin{itemize}
\item  3. $(M,\gamma )$ is of negative Yamabe class, we take $r=-1.$. 
\begin{equation}
h(x,y)\equiv b(x)y^{\frac{4n}{n-2}}-y^{\frac{4(n-1)}{n-2}}-q_{1}(x)y^{\frac{%
2(n-1)}{n-2}}-a(x)
\end{equation}
\begin{equation}
h_{y}^{\prime }\equiv \frac{2}{n-2}\tilde{h}(x,y)y^{\frac{n}{n-2}},\text{ \ }%
\tilde{h}(x,y)\equiv 2nby^{^{\frac{2n+2}{n-2}}}-2(n-1)y^{\frac{2n-2}{n-2}%
}-(n-1)q_{1}.
\end{equation}
\begin{equation}
\tilde{h}_{y}^{\prime }\equiv \frac{4}{n-2}\{n(n+1)by^{^{\frac{4}{n-2}%
}}-(n-1)^{2}\}y^{\frac{n}{n-2}}.
\end{equation}
If $b(x)>0$ then $\tilde{h}(x,.)$ starts from a non positive value,
decreases until $y=Y_{0}(x),$%
\begin{equation}
Y_{0}(x)=\{\frac{(n-1)^{2}}{n(n+1)b(x)}\}^{\frac{n-2}{4}},
\end{equation}
increases afterwards up to +$\infty ,$ it has therefore one positive zero $%
Y(x)>Y_{0}(x).$ Hence $h(x,.)$ starts from a non positive value, decreases
until $y=Y(x)$ then increases up to $+\infty .$ We see that if $Inf$ $b>0$
on $M$ one can always find $\ell $ and $m$ satisfying 4.4 by choosing $\ell $
and $m$ such that: 
\begin{equation}
0<\ell \leq \underset{M}{Inf}Y(x),\text{ \ \ \ }m\geq \underset{M}{Sup}Y(x).
\end{equation}
\end{itemize}
\end{proof}

Sub and supersolutions do not need to be constants: we will obtain the
following theorem by looking for non constant sub or supersolutions. It
replaces inequalities to be satisfed on $M$ by inequalities on an open set
of $M.$

\begin{corollary}
The theorem 4.10 holds if

1. The hypothesis $\underset{M}{Inf}(a+q_{1})>0$ is replaced by $%
a+q_{1}\not\equiv 0$ for the positive or zero Yamabe classes.

2.\ The hypothesis $\underset{M}{Inf}\tau ^{2}>0$ is replaced by:

a. $\tau ^{2}\not\equiv 0$ for the zero Yamabe class.

b. $Inf$ $\tau ^{2}>0$ on a sufficiently large subset of $M$ for the
negative Yamabe class.
\end{corollary}

\begin{proof}
1. We show as in Isenberg 1987 that the condition $\underset{M}{Inf}%
(a+q_{1})>0$ is superfluous for the construction of a non constant $\varphi
_{-}$, by constructing a constant subsolution for a conformally transformed
equation. Suppose that $a+q_{1}\not\equiv 0,$ i.e. there is an open subset $%
U\subset M$ where $\underset{U}{Inf}(a+q_{1})>0.$ The complementary set $M-U$
is contained in a proper open subset $V$ which we can take with smooth
boundary $\partial V$. We can construct a metric $\gamma $' conformal to $%
\gamma $ whose scalar curvature is stricly negative in $V$, as follows. Set $%
\gamma ^{\prime }=\theta ^{4/(n-2)}\gamma ,$ the formula \ for the conformal
transformation of the scalar curvature gives: 
\begin{equation}
r(\gamma ^{\prime })\equiv \theta ^{-\frac{n+2}{n-2}}(-\Delta _{\gamma
}\theta +r(\gamma )\theta ).
\end{equation}
Take for $\theta \in W_{3}^{p}$ the unique strictly positive solution in $V$
of the linear equation and boundary condition, with $k$ some positive
constant, 
\begin{equation}
\Delta _{\gamma }\theta -k\theta =0,\text{ \ \ }\theta \mid _{\partial V}=1.
\end{equation}
These two equations imply that: 
\begin{equation*}
r(\gamma ^{\prime })=\theta ^{-\frac{n+2}{n-2}}(r(\gamma )-k).
\end{equation*}
We take $k>\underset{M}{Sup}r(\gamma )$ we have then $r(\gamma ^{\prime
})<0, $ with $\underset{V}{Inf}|r(\gamma ^{\prime })|>0,$ in $V$. We choose
on $M$ a smooth function $\theta >0$ equal on $V$ to the previously
determined $\theta $. The function $\varphi ^{\prime }=\theta ^{-1}\varphi $
satisfies the following equation, where $a^{\prime }=a\theta ^{-\frac{4n}{n-2%
}},$ $q_{1}^{\prime }=q_{1}\theta ^{-\frac{4n}{n-2}},$ hence \ $\underset{M-V%
}{\inf }(a^{\prime }+q_{1}^{\prime })>0$: 
\begin{equation*}
\Delta _{\gamma ^{\prime }}\varphi ^{\prime }-r(\gamma ^{\prime })\varphi
^{\prime }+a^{\prime }\varphi ^{\prime -\frac{3n-2}{n-2}}+q_{1}^{\prime
}\varphi ^{\prime -\frac{n}{n-2}}-b\varphi ^{\prime \frac{n+2}{n-2}}=0.
\end{equation*}
The number $\ell ^{\prime }>0$ will be a subsolution of this equation if it
is such that (we impose $\ell ^{\prime }\leq 1,$ which is no restriction, to
simplify the writing of the second inequality): 
\begin{equation}
0<l^{\prime }\leq Min\{1,\frac{\underset{V}{Inf}|r(\gamma ^{\prime })|}{%
\underset{V}{Sup}b},\frac{\underset{M-V}{Inf}(a^{\prime }+q_{1}^{\prime })}{%
\underset{M-V}{Sup}(b+|r(\gamma ^{\prime }|}\}.
\end{equation}
The function $\varphi _{-}\equiv \theta \ell ^{\prime }>0$ is a subsolution
of the original Lichnerowicz equation. It is always possible to choose the
constant supersolution $m$ large enough such that $\varphi _{-}\leq m$ on $%
M. $

2. If we know only that $\underset{U_{1}}{Inf}b>0,$ $U_{1}$ a subset of $M,$
the problem is with the supersolutions. We try to conformally transform the
metric $\gamma $ to a metric with positive scalar curvature in $V_{1},$ with 
$M-U_{1}\subset V_{1},$ by considering now the equation in $V_{1}$ and
boundary condition, with $k$ some positive number, 
\begin{equation}
\Delta _{\gamma }\theta +k\theta =0,\text{ \ \ }\theta \mid _{\partial
V_{1}}=1.
\end{equation}
This equation implies that: 
\begin{equation*}
r(\gamma _{1})\equiv \theta ^{-\frac{n+2}{n-2}}(r(\gamma )+k),\text{ \ if \
\ }\gamma _{1}=\theta ^{4/(n-2)}\gamma .
\end{equation*}
with $r(\gamma _{1})>0$ if 
\begin{equation}
k>|r(\gamma )|
\end{equation}
The problem 4.34 is of Fredholm type: it has a solution if the homogeneous
problem, i.e. the equation 4.34 but with boundary condition $\theta
_{\partial V_{1}}=0,$ has for unique solution $0,$ which is the case for $%
k>0,$ if $k$ is smaller that the first eigenvalue. In this case the solution 
$\theta $ is positive. The conformally transformed equation (we extend $%
\theta $ to $M,$ as above) has constant supersolutions, all numbers $m_{1}$
such that 
\begin{equation}
m_{1}\geq Max\{1,\frac{\underset{U_{1}}{Sup}(|r(\gamma _{1})|+a_{1}+q_{1})}{%
\underset{U_{1}}{Inf}b},\frac{\underset{V_{1}}{Sup}(a_{1}+q_{1})}{\underset{%
V_{1}}{Inf}\theta _{1}^{-(n+2)/(n-2)}r(\gamma _{1})}\},\text{ \ \ }
\end{equation}
A supersolution $\varphi _{+}$ of the original equation, arbitrarily large
(but bounded), such that $\varphi _{-}\leq \varphi _{+},$ is deduced from
the inverse conformal transformation.

a. If $r(\gamma )=0$ the condition 4.35 is satisfied by any positive $k.$

b. If $r(\gamma )=-1$, the condition 4.35 is verified with a number $k$
smaller than the first eigenvalue $\lambda $ of the homogeneous problem . It
holds that $\lambda =C_{F}^{-1}$ where $C_{F}$ is the Friedrichs constant of
the domain $(V_{1},\gamma ),$ smallest number such that: 
\begin{equation}
\int_{V_{1}}|u|^{2}\mu _{\gamma }\leq C_{F}\int_{V_{1}}|Du|^{2}\mu _{\gamma
},\text{ \ for all \ \ }u\in \mathcal{D(}V_{1}).
\end{equation}
\end{proof}

\textbf{Conclusion. }Existence and non existence results cover all cases
when $b$, that is $\tau ,$ is a constant, since then either $\underset{M}{Inf%
}b>0$ or $b\equiv 0.$ A gap remains when $\tau $ is not constant and $%
(M,\gamma )$ is in the negative Yamabe class.

\textbf{Remark.} The method can be applied in the presence of \textbf{%
unscaled sources}, with practically no change if $b-q_{2}\geq 0.$ If $%
b-q_{2}<0$ there can be solutions only if $\gamma $ is in the positive
Yamabe class (see a construction of sub and supersolutions in the case $n=3$
in CB\ 1972). In the general case the discussion is more involved, but seems
to offer no conceptual difficulty.

\subsection{ Scaled sources in $L^{p}$.}

Though discontinuous energy sources are allowed by our hypothesis $q_{1}\in
L^{\infty }$, the following theorem, deduced from theorem 8.2, is useful in
the presence of less regular sources and for the coupling with the momentum
constraint with low regularity sources.

\begin{theorem}
If $(M,\gamma )$ is in the positive Yamabe class the theorem 4.10 and
corollary 4.11 hold with hypothesis on $a$ and $q_{1}$ weakened to $%
a,q_{1}\in L^{p}.$

The same result holds for the zero Yamabe class under the additional
assumption $b\not\equiv 0.$
\end{theorem}

\begin{proof}
The problem in the case where $a+q_{1}$ is not bounded above is in the
construction of a constant supersolution $m.$ To construct a non constant
supersolution, in the positive Yamabe case, we proceed as in Moncrief 1986
and CB-Moncrief 1994 (case $n=2$). For simplicity we write up the physical
case $n=3.$ We \ denote by $\varphi _{0}^{4}\equiv y_{0}$ the positive
number solution of the equation 
\begin{equation}
\underline{b}y_{0}^{3}+\underline{r}y_{0}^{2}-\underline{q}_{1}y_{0}-%
\underline{a}=0.
\end{equation}
where \underline{$f$} denotes the mean value of $f$ on $(M,\gamma ):$%
\begin{equation}
\underline{f}\equiv \frac{1}{Vol(M,\gamma )}\int_{M}f\mu _{\gamma }.
\end{equation}
Such a number exists if $\underline{r}>0$ or $\underline{r}=0$ and 
\underline{$b$}$>0.$

We define one function $v\in W_{2}^{p},$ with mean value zero on $M,$ by
solving the linear equation 
\begin{equation}
\Delta _{\gamma }v=r\varphi _{0}-a\varphi _{0}^{-7}-q_{1}\varphi
_{0}^{-3}+b\varphi _{0}^{5}
\end{equation}
The function 
\begin{equation}
\varphi _{+}\equiv \varphi _{0}+v-Infv\geq \varphi _{0},\text{ \ \ }\Delta
_{\gamma }\varphi _{+}\equiv \Delta _{\gamma }v,
\end{equation}
is a supersolution because it holds that: 
\begin{equation}
\Delta _{\gamma }\varphi _{+}-f(.,\varphi _{+})=r(\varphi _{0}-\varphi
_{+})-a(\varphi _{0}^{-7}-\varphi _{+}^{-7})-q_{1}(\varphi _{0}^{-3}-\varphi
_{+}^{-3})+b(\varphi _{0}^{5}-\varphi _{+}^{5})
\end{equation}
hence if $r\geq 0$ 
\begin{equation}
\Delta _{\gamma }\varphi _{+}-f(.,\varphi _{+})\leq 0.
\end{equation}
Since $\varphi _{+}\geq \varphi _{0}>0$ we can choose a subsolution $\ell >0$
such that $0<\ell \leq \varphi _{+}$ and apply the theorem 8.2.
\end{proof}

When $n=3$ our result includes the case $p=2,$ i.e. $\gamma ,$ $\varphi \in
H_{2}.$

\section{Coupled system, $\protect\tau =$constant, scaled sources.\ }

When $\tau $ is a constant the momentum constraint (scaled sources) does not
depend on $\varphi .$ We can therefore solve this linear system for $X$ and
insert afterwards the result in the semilinear equation for $\varphi .$ To
solve this Lichnerowicz equation we have supposed that $\gamma \in
M_{2}^{p}, $ $p>\frac{n}{2},$ and the coefficients $a,b,q_{1}\in L^{\infty }$%
, except $a $ and $q_{1}$ in the positive Yamabe case. While $b$ and $q_{1}$
are given quantities, the function $a$ depends on the given $\gamma $ and $%
B, $ and also on the unknown $X,$ solution of the momentum constraint. We
leave to the reader formulations of variants of the following existence
theorem.

\begin{theorem}
Let $\gamma \in M_{2}^{p}.$ Let $q_{1}$ and $J_{1},$ $L^{2}$ orthogonal to
the conformal Kiling fields of $(M,\gamma ),$ be given as well as $B\in
W_{1}^{p},$ and $\tau =$constant. Then the LCBY system with scaled sources
admits a solution $X,\varphi \in W_{2}^{p},$ $\varphi >0,$ in the following
cases

1. ($M,\gamma )$ is in the positive Yamabe class$,$ $p>\frac{n}{2},$ $%
q_{1},J_{1}\in L^{p},$ $B\in W_{1}^{p}$ and $|J|_{1}+q_{1}\not\equiv 0$ on $%
M $.

2. ($M,\gamma )$ is in the zero Yamabe class$,$ $p>\frac{n}{2},$ $q_{1},$ $%
J_{1}\in L^{p},$ $\tau \not=0$ and $|J|_{1}+q_{1}\not\equiv 0$ on $M.$

3.($M,\gamma )$ is in the \ negative Yamabe class $p>n,$ $q_{1}\in L^{\infty
},$ $J_{1}\in L^{p},$ $B\in W_{1}^{p},$ $\tau \not=0$.

The solution is unique, except if $(M,\gamma )$ is in the zero Yamabe class
and $|J|_{1}+|B|+q_{1}+\tau ^{2}\equiv 0.$
\end{theorem}

\begin{proof}
Under the hypothesis the momentum constraint 3.1 has a solution $X\in
W_{2}^{p},$ since the vector $F$ given by 3.2, here independent of $\varphi $
and $J_{2},$ is in $L^{p}$ and is $L^{2}$ orthogonal to the space of
conformal Killing (CK) vector fields of ($M,\gamma ).$ The tensor \ $A\equiv 
\mathcal{L}_{\gamma ,conf}X+B$ is then in $W_{1}^{p}$. The function $a\equiv
A.A$ is in $L^{p}$ as soon as $p>\frac{n}{2}.$

1.\ and 2 The existence theorem 4.12, positive or zero Yamabe class, applies
if $a+q_{1}\not\equiv 0.$ Since 
\begin{equation}
D_{i}A^{ij}=J_{1}^{j}
\end{equation}
we have $A\not\equiv 0,$ hence $a\not\equiv 0,$ if $J_{1}\not\equiv 0.$
Therefore $|J|_{1}+q_{1}\not\equiv 0$ implies $a+q_{1}\not\equiv 0.$

3.\ If $p>n$ the tensor $A$ is in $C^{0},$ hence the function $a$ is also in 
$C^{0}\subset L^{\infty }.$ We apply the theorem 4.10 and its corollary 4.11.
\end{proof}

\section{Solutions with non constant TrK, or (and) unscaled sources.}

We will show that there exists a whole neighbourhood of low regularity
solutions with non constant\footnote{{\footnotesize There are cosmological
spacetimes which do not admit surfaces of constant }$\tau ${\footnotesize \
(Bartnik 1988)}} $\tau $ and (or) non zero unscaled sources near such a
solution with constant $\tau $ and zero unscaled sources.

\begin{lemma}
The mapping defined by: 
\begin{equation}
\Phi :(x,y)\mapsto \Phi (x,y)\equiv (\mathcal{H(}x,y),\mathcal{M}(x,y)),
\end{equation}
\begin{equation}
\text{\ }x\equiv (\gamma ,\tau ,B,q_{1},q_{2},J_{1},J_{2}),\text{ \ }y\equiv
(\varphi ,X)
\end{equation}
\begin{equation}
\mathcal{H}(x,y)\equiv \Delta _{\gamma }\varphi -r(\gamma )\varphi
+a(B,X)\varphi ^{-\frac{3n-2}{n-2}}+q_{1}\varphi ^{-\frac{n}{n-2}}-(q_{2}-%
\frac{n-2}{4n}\tau ^{2})\varphi ^{\frac{n+2}{n-2}}
\end{equation}
\begin{equation}
\mathcal{M}^{i}(x,y)\equiv (\Delta _{\gamma ,conf}X)^{i}-\{\frac{n-1}{n}%
\varphi ^{2n/(n-2)}\partial ^{i}\tau +J_{1}^{i}+J_{2}^{i}\varphi
^{2(n+2)/(n-2)}-D_{j}B^{ij}\}
\end{equation}
is, if $p>\frac{n}{2},$ a differentiable map into the Banach space $\mathbf{B%
}_{0}\equiv L^{p}\times ^{1}\otimes L^{p}$ from the open set $\Omega \equiv
\Omega _{1}\times \Omega _{2}$ of the Banach space $\mathbf{B\equiv B}%
_{1}\times \mathbf{B}_{2}$, defined by\footnote{{\footnotesize To help the
reader follow our reasoning we denote by }$^{m}\otimes W_{s}^{p}$ 
{\footnotesize the space of }$W_{s}^{p}$ {\footnotesize \ m-\ tensor fields.}%
} 
\begin{equation}
\Omega _{1}\equiv \mathbf{B}_{1}\cap \{\gamma \in M_{2}^{p}\},\text{ \ \ }%
\Omega _{2}\equiv \mathbf{B}_{2}\cap \{\varphi >0\}
\end{equation}
\begin{equation}
\mathbf{B}_{1}\equiv (W_{2}^{p}\times W_{1}^{p}\times (^{2}\otimes
W_{1}^{p})\times L^{p}\times L^{p}\times ^{1}\otimes L^{p}\times ^{1}\otimes
L^{p}),
\end{equation}
\begin{equation}
\mathbf{B}_{2}\equiv W_{2}^{p}\times (^{1}\otimes W_{2}^{p}).
\end{equation}
\end{lemma}

\begin{proof}
The mapping $\Phi $ is continuous from $\Omega $ into $L^{p}\times
^{1}\otimes L^{p}$ because $\varphi \in W_{2}^{p}\subset C^{0},$ $\varphi
>0, $ and $DB,\partial \tau ,a,q_{1},q_{2},J_{1},J_{2}\in L^{p}$ as well as $%
\tau ^{2}$ and $r\equiv r(\gamma )$ when $p>\frac{n}{2},$ $.$ It is
differentiable on $\Omega $ because the linear mapping 
\begin{equation}
\delta \Phi _{x,y}:(\delta x,\delta y)\mapsto \delta \Phi _{x,y}(\delta
x,\delta y)\equiv (\delta \mathcal{H}_{x,y}(\delta x,\delta y),\delta 
\mathcal{M}_{x,y}(\delta x,\delta y))
\end{equation}
is a mapping from the Banach space $\mathbf{B}$ into the Banach space $%
L^{p}\times ^{1}\otimes L^{p}$ for each $(x,y)\in \Omega .$ Indeed this
linear mapping is given by: 
\begin{equation*}
\delta \mathcal{H}_{x,y}(\delta x,\delta y)\equiv \Delta _{\gamma }\delta
\varphi -\{r+\frac{3n-2}{n-2}a\varphi ^{-\frac{2n}{n-2}}+\frac{n}{n-2}%
q_{1}\varphi ^{-2}+\frac{n+2}{n-2}(b-q_{2})\varphi ^{\frac{4}{n-2}}\}\delta
\varphi
\end{equation*}
\begin{equation}
+\delta \gamma .D^{2}\varphi +\delta (\Gamma (\gamma )).D\varphi -\delta
r\varphi +\delta a\varphi ^{-\frac{3n-2}{n-2}}+\delta q_{1}\varphi ^{-\frac{n%
}{n-2}}+(\delta q_{2}-\delta b)\varphi ^{\frac{n+2}{n-2}}
\end{equation}
and 
\begin{equation*}
\delta \mathcal{M}_{x,y}(\delta x,\delta y)\equiv (\Delta _{\gamma
,conf}\delta X)^{i}-(\frac{2(n-1)}{n-2}\varphi ^{\frac{n+2}{n-2}}\partial
^{i}\tau +\frac{2(n+2)}{n-2}\varphi ^{\frac{n+6}{n-2}}J_{2}^{i})\delta
\varphi
\end{equation*}
\begin{equation}
+D_{j}\delta B^{ij}+(\delta (\Gamma (\gamma )).B)^{i}-\{\frac{n-1}{n}\varphi
^{\frac{2n}{n-2}}\partial ^{i}\delta \tau +\varphi ^{\frac{2(n+2)}{n-2}%
}\delta J_{2}^{i}+\delta J_{1}^{i}\}.
\end{equation}
Embeddings and multiplication properties give as in previous sections the
announced result.
\end{proof}

\begin{theorem}
Let ($\bar{g},\bar{K}),$ $\bar{K}=\bar{B}+\mathcal{L}_{\bar{g},conf}\bar{X}+%
\frac{\bar{g}}{n}\bar{\tau}$ be a solution of the Einstein constraints on
the compact n-manifold $M,$ with TrK$\equiv \bar{\tau}$ a constant, $\bar{g}%
\in M_{2}^{p},$ $\bar{X}\in W_{2}^{p},$ $\bar{B}\in W_{1}^{p},$ $p>\frac{n}{2%
}$ and scaled sources $\bar{q}_{1},$ $\bar{J}_{1}\in L^{p}$ and no unscaled
sources. Suppose $(M,\bar{g})$ admits no conformal Killing field. There is a
neighbourhood of $\bar{\tau},\bar{B},$ $\bar{q}_{1},$ $\bar{q}_{2}=0,$ $\bar{%
J}_{1},$ $\bar{J}_{2}=0$ in $\mathbf{B}_{1}$ such that the Einstein
constraints with non constant $\tau $ and (or) unscaled sources admit a
solution ($g,K)$ in a $W_{2}^{p}\times W_{1}^{p}$ neighbourhood of ($\bar{g},%
\bar{K}).$
\end{theorem}

\begin{proof}
We have just proved that the mapping $\Phi $ defined by the left hand side
of the LCBY equations is a differentiable map into a Banach space $\mathbf{B}%
_{2}$. Saying that the function $\bar{\varphi}=1$ together with $\bar{X}\in
W_{2}^{p}$ is a solution of the LCBY system with data $\bar{\gamma}\equiv 
\bar{g}\in M_{2}^{p},$ $\bar{\tau}$ a constant$,$ $\bar{q}_{1}$ $\in L^{p},,%
\bar{q}_{2}=0$, $\bar{B}\in W_{1}^{p},$ $\bar{J}_{1}\in L^{p},$ $\bar{J}%
_{2}=0$ is to say that, with the notations of the previous lemma: 
\begin{equation}
\Phi (\bar{x},\bar{y})=0
\end{equation}
$.$By the implicit function theorem\footnote{{\footnotesize See for instance
CB and DeWitt I \ p.91.}} there exists of a solution $y\equiv $($\varphi
,X)\in \Omega _{2},$ i.e. $\varphi \in W_{2}^{p},$ $\varphi >0,$ $X\in
W_{2}^{p},$ when $x\equiv \{\gamma \in M_{2}^{p},$ $\tau \in W_{1}^{p},$ $%
B\in W_{1}^{p},$ $q_{1},$ $q_{2},$ $J_{1},$ $J_{2}\in L^{p}\}$ is in a
sufficiently small neighbourhood in the Banach space $\mathbf{B}$\ of $\bar{x%
}$, under the condition that the partial derivative of $\Phi $ with respect
to $y$ at $(\bar{x},\bar{y})$ is an isomorphism of Banach spaces, that is if
the linear system 
\begin{equation}
(\Delta _{\bar{g},conf}\delta X)^{i}=Y
\end{equation}
and (recall that $\bar{\varphi}=1)$%
\begin{equation}
\Delta _{\bar{g}}\delta \varphi -\{\bar{r}+\frac{3n-2}{n-2}\bar{a}+\frac{n}{%
n-2}\bar{q}_{1}+\frac{n+2}{n-2}\bar{b}\}\delta \varphi =\psi
\end{equation}
has one and only one solution $\delta X\in ^{1}\otimes W_{2}^{p},$ $\delta
\varphi \in W_{2}^{p}$ for each $Y\in ^{1}\otimes L^{p},$ $\psi \in L^{p}.$
This result holds for 6.11 if $\bar{g}$ admits no conformal Killing field.
It holds for 6.12 if the coefficient of $\delta \varphi ,$ which belongs to $%
L^{p},$ $p>\frac{n}{2}$ by hypothesis, is positive or zero, and not
identically zero. It is obviously so when $\bar{r}>0.$ To treat general
values of $\bar{r}$\ we remark, as O'Murchada and York 1974 in their study
of linearisation stability, that since ($\bar{g},\bar{K})$ satisfies the
hamiltonian constraint, hence the Lichnerowicz equation with $\bar{\varphi}%
=1,$ it holds that: 
\begin{equation}
-\bar{r}+\bar{a}+\bar{q}_{1}-\bar{b}=0.
\end{equation}
The linear equation 6.12 reads therefore 
\begin{equation}
\Delta _{\bar{g}}\delta \varphi -\{2\bar{a}+\frac{2(n-1)}{n-2}\bar{q}_{1}+%
\frac{4}{n-2}\bar{b}\}\delta \varphi =\psi
\end{equation}
The coefficient of the lower order term $\delta \varphi $ is positive or
zero, and not identically zero, if $\bar{a},$ $\bar{q}_{1},$ and $\bar{b}$
are not all identically zero.
\end{proof}

The result does not extend in a straightforward way\footnote{{\footnotesize %
An analogous problem occurs in linearization stability (see Moncrief 1975)}}
to the case where $\bar{g}$ admits conformal Killing fields because $%
\mathcal{M}$ does not takes its values in the space of vectorfields
orthogonal to such fields.

\section{Appendix A. Second order linear elliptic systems.}

\subsection{Linear systems}

Let $(M,e)$ be a smooth compact orientable riemannian manifold. Denote by $%
\partial $ the covariant derivative in the metric $e$. A second order linear
differential operator from sections $u$ of a tensor bundle $E$ over $(M,e)$
into sections of another such bundle $F$ reads 
\begin{equation}
Lu\equiv \sum_{k=0}^{2}a_{k}\partial ^{k}u
\end{equation}
with $a_{k}$ a linear map from tensor fields to tensor fields given also by
tensor fields over $M$. In local coordinates it reads: 
\begin{equation}
(Lu)^{A}\equiv a_{2,B}^{A,ij}\partial _{ij}^{2}u^{B}+a_{1,B}^{A,i}\partial
_{i}u^{B}+a_{0,B}^{A}u^{B}
\end{equation}

The principal symbol of the operator $L$ at a point $x\in M$, for a covector 
$\xi $ at $x$, is the linear map from $E_{x}$ to $F_{x}$ determined by the
contraction of $a_{2}$ with $(\otimes \xi )^{2}$, represented in coordinates
by the matrix 
\begin{equation}
M_{B}^{A}(\xi )\equiv a_{2,B}^{A,ij}\xi _{i}\xi _{j}
\end{equation}
The operator is said to be elliptic if for each $x\in M$ and $\xi \in
T_{x}^{\ast }M$ its principal symbol is an isomorphism from $E_{x}$ onto $%
F_{x}$ for all $\xi \neq 0,$ i.e. the determinant of the matrix $M_{B\text{ }%
}^{A}$ does not vanish. If there is a covering of $M$ such that this
determinant is uniformly bounded away from zero the system is said to be
uniformly elliptic and the bound is called the ellipticity constant. The
Laplace operator $\Delta _{\gamma }$ in a riemannian metric $\gamma $ on $M$
acting on scalar functions $u$ has principal symbol $\gamma ^{ij}\xi _{i}\xi
_{j},$ it is elliptic.

The Sobolev spaces $W_{s}^{p}$ on $(M,e)$ are defined as closures of spaces
of smooth tensor fields in the norm 
\begin{equation}
||f||_{W_{s}^{p}}\equiv \{\int_{M}\sum_{0\leq k\leq s}|\partial
^{k}f|^{p}\mu _{e}\}^{\frac{1}{p}},
\end{equation}
where $\partial $, $|.|$ and $\mu _{e}$ denote the covariant derivative and
the pointwise norm in the metric $e.$

We denote by $C$ any positive number depending only on $(M,e):$ volume,
Sobolev constant...

A metric $\gamma $ is said to belong to $M_{\sigma }^{p}$ if it is properly
riemannian and $\gamma \in W_{\sigma }^{p}.$ We will always suppose that $%
\sigma >\frac{n}{p},$ then $\gamma \in M_{\sigma }^{p}$ is continuous on $M$
and $M_{\sigma }^{p}$ is an open subset of $W_{\sigma }^{p}.$ The
contrevariant associate $\gamma ^{\#}$ is also continuous and the volume
elements $\mu _{\gamma }$ and $\mu _{e}$ are equivalent.

We recall the following theorem.

\begin{theorem}
(Douglis and Nirenberg) Let $\Omega $ be a bounded open set of $R^{n},$ let $%
\tilde{L}$ be an homogeneous elliptic operator: 
\begin{equation}
\tilde{L}u\equiv a_{m}\tilde{\partial}^{m}u
\end{equation}
where $\tilde{\partial}$ is the usual partial derivative and $a_{m}$ is
continuous and bounded on $\Omega ,$ then the following estimate holds, for
any $p>1$: 
\begin{equation}
\Vert u\Vert _{\tilde{W}_{m}^{p}}\leq C_{a_{m}}\{\Vert \tilde{L}u\Vert _{%
\tilde{L}^{p}}+||u||_{\tilde{L}^{p}}\}
\end{equation}
where $\tilde{W}_{s}^{p}$ denotes Sobolev spaces defined by the euclidean
metric on $\Omega .$ The number $C_{a_{m}}$ depends only on $\Omega $ the
ellipticity constant, the $C^{0}$ norm of $a_{m}$ and its modulus of
continuity.
\end{theorem}

We will deduce from this theorem the following one\footnote{{\footnotesize %
Lowering the regularity of the coefficients of the theorem proved in CB-Ch,
it can be extended to higher order elliptic operators and possibly lower
values of p for the solution, but we will need the restriction p\TEXTsymbol{>%
}n/2 when working with the non linear Lichnerowicz equation and treat here
only this case for simplicity.}}.

\begin{theorem}
Let $(M,e)$ be a smooth compact orientable Riemannian manifold. Let 
\begin{equation}
Lu\equiv \sum_{k=0}^{2}a_{k}\partial ^{k}u
\end{equation}
be a second order elliptic operator on $(M,e)$. Suppose the coefficients of $%
L$ are such that: 
\begin{equation*}
a_{2}\in W_{2}^{p},\text{ }a_{1}\in W_{1}^{p},\text{\ }a_{0}\in L^{p},\text{
\ \ }p>\frac{n}{2}.
\end{equation*}
Then

1. The operator $L$ is a continuous mapping $W_{2}^{p}\rightarrow L^{p}.$

2. The following estimate holds: 
\begin{equation}
\Vert u\Vert _{W_{2}^{p}}\leq C_{L}\{\Vert Lu\Vert _{L^{p}}+||u||_{L^{1}}\}.
\end{equation}
The number $C_{L}$ depends only on the norms of the $a_{k}^{\prime }s$ in
their respective spaces and the ellipticity constant of $a_{2}$.
\end{theorem}

\begin{corollary}
If morevover 
\begin{equation*}
a_{2}\in W_{2+s}^{p},\text{ }a_{1}\in W_{1+s}^{p},\text{\ }a_{0}\in
W_{s}^{p},
\end{equation*}
then $L$ is a continuous mapping $W_{s+2}^{p}\rightarrow W_{s}^{p}$ and for
all $u\in W_{s+2}^{p}$ it holds that: 
\begin{equation}
\Vert u\Vert _{W_{s+2}^{p}}\leq C_{L}\{\Vert Lu\Vert
_{W_{s}^{p}}+||u||_{L^{1}}\}.
\end{equation}
with $C_{L}$ a number depending only on the norms of the $a_{k}^{\prime }s$
and the ellipticity constant of $a_{2}$.
\end{corollary}

\begin{proof}
By the Sobolev embedding theorem we know that $a_{2}$ belongs to the
H\"{o}lder space $C^{0,\alpha \text{ }}$since $W_{2}^{p}\subset C^{0,\alpha
} $ \ if \ $p>\frac{n}{2}.$ Therefore $a_{2}\partial ^{2}u\in L^{p}$ if $%
u\in W_{2}^{p}.$

If $a_{1}\in W_{1}^{p}$ then $a_{1}\partial u\in L^{p}$ due to the
continuous multiplication $W_{1}^{p}\times W_{1}^{p}\rightarrow L^{p}$ when $%
p>\frac{n}{2}$ and it holds that 
\begin{equation}
||a_{1}\partial u||_{L^{p}}\leq C||a_{1}||_{W_{1}^{p}}||\partial
u||_{W_{1}^{p}}
\end{equation}

If $a_{0}\in L^{p}$ then $a_{0}u\in L^{p}$ whatever be $u$ since $u\in
C^{0,\alpha }.$

To prove the inequality 7.8 using the Douglis - Nirenberg theorem we proceed
in two steps.

a. We treat the principal part by considering a covering of $M$ by a finite
number of charts with bounded domains $\Omega _{i},$ and a partition of
unity, functions $0\leq \phi _{i}\leq 1$ with compact support in $\Omega
_{i} $ such that $\sum_{i}\phi _{i}=1.$ We have $u=\sum u_{i},$ $u_{i}\equiv
u\phi _{i}.$ In any of the charts $e$ is uniformly equivalent to the
euclidean metric; we denote by $\tilde{\partial}$ the derivative in the
euclidean metric (i.e. the usual partial derivative), by $\tilde{W}_{s}^{p}$
the associated Sobolev norm. For fields with support in a chart the norms $%
W_{s}^{p}$ and $\tilde{W}_{s}^{p}$ are equivalent, they are uniformy
equivalent for the various charts, since there is a finite number of them.
The Douglis - Nirenberg theorem gives then the existence of a constant $%
C_{i},$ depending only on $\Omega _{i}$ and on $a_{2}$ through its $%
C^{0,\alpha }$ norm and the ellipticity constant of its representatives,
such that 
\begin{equation}
\Vert u_{i}\Vert _{W_{2}^{p}}\leq C_{i,a_{2}}\{\Vert \tilde{L}u_{i}\Vert
_{L^{p}}+||u||_{L^{p}}\},\text{ \ \ \ }\tilde{L}u_{i}\equiv a_{2}\tilde{%
\partial}^{2}u_{i}.
\end{equation}
The subadditivity of norms together with this inequality imply that, with $%
C_{a_{2}}$ the maximum of the $C_{i}^{\prime }s,$%
\begin{equation}
||u||_{W_{2}^{p}}\leq \sum_{i}||u_{i}||_{W_{2}^{p}}\leq
C_{a_{2}}\sum_{i}\{\Vert \tilde{L}u_{i}\Vert _{L^{p}}+||u||_{L^{p}}\}.
\end{equation}
On the other hand, we have an identity of the form, with $S(i)$ smooth
coefficients linear in the connection of $e:$ 
\begin{equation}
\tilde{L}u_{i}\equiv a_{2}\tilde{\partial}^{2}u_{i}\equiv a_{2}\{\partial
^{2}u_{i}+S(i)\partial u_{i}\}
\end{equation}
Using the Leibniz rule we obtain that 
\begin{equation}
\partial u_{i}=u\partial \phi _{i}+\partial u\phi _{i}\text{, \ \ \ \ }%
\partial ^{2}u_{i}=\partial ^{2}u\phi _{i}+2\partial u\partial \phi
_{i}+u\partial ^{2}\phi _{i}.
\end{equation}
Since 0$\leq \phi _{i}\leq 1$ we deduce from 7.14 that 
\begin{equation}
|a_{2}\partial ^{2}u_{i}|\leq |a_{2}\partial ^{2}u|+|a_{2}(2\partial
u\partial \phi _{i}+u\partial ^{2}\phi _{i})|.
\end{equation}
Combining previous inequalities we find that there exists a number $C$
depending only on ($M,e),$ such that 
\begin{equation}
||\tilde{L}u_{i}||_{L^{p}}\leq C\{||a_{2}\partial
^{2}u||_{L^{p}}+||a_{2}||_{C^{0}}||u||_{W_{1}^{p}}\}.
\end{equation}
The inequality 7.12 implies then that: 
\begin{equation}
||u||_{W_{2}^{p}}\leq C_{a_{2}}\{||a_{2}\partial
^{2}u||_{L^{p}}+||u||_{L^{p}}\}.
\end{equation}

b. We now consider the general linear operator $L$, with coefficients
satisfying the given hypotheses: 
\begin{equation}
Lu\equiv a_{2}\partial ^{2}u+a_{1}\partial u+a_{0}u.
\end{equation}
Since $a_{2}\in W_{2}^{p},$ with $p>\frac{n}{2},$ it belongs also to a
H\"{o}lder space $C^{0,\alpha },$ $0<\alpha \leq 2-\frac{n}{p},$ and it
holds that, with $C$ a Sobolev constant of ($M,e)$, 
\begin{equation}
||a_{2}||_{C^{0,\alpha }}\leq C||a_{2}||_{W_{2}^{p}}.
\end{equation}

The inequality 7.17\ implies that 
\begin{equation}
||u||_{W_{2}^{p}}\leq C_{a_{2}}\{||Lu||_{L^{p}}+||a_{1}\partial
u+a_{0}u||_{L^{p}}+||u||_{W_{1}^{p}}\}.
\end{equation}

To estimate $||a_{1}\partial u||_{L^{p}}$ we use the H\"{o}lder inequality 
\begin{equation}
||a_{1}\partial u||_{L^{p}}\leq ||a_{1}||_{L^{p_{1}}}||\partial u||_{L^{q}},%
\text{ \ \ }\frac{1}{p_{1}}+\frac{1}{q}=\frac{1}{p}.
\end{equation}
The Sobolev embedding theorem gives that 
\begin{equation}
||a_{1}||_{L^{p_{1}}}\leq C||a_{1}||_{W_{1}^{p}},\text{ \ }p_{1}=\frac{np}{%
n-p}.
\end{equation}
Taking this value of $p_{1}$ gives that 
\begin{equation}
\frac{1}{q}=\frac{1}{p}-\frac{1}{p_{1}}=\frac{1}{n}.
\end{equation}
and 
\begin{equation}
||a_{1}\partial u||_{L^{p}}\leq C||a_{1}||_{W_{1}^{p}}||\partial u||_{L^{n}}.
\end{equation}
On the other hand: 
\begin{equation}
||a_{0}u||_{L^{p}}\leq ||a_{0}||_{L^{p}}||u||_{L^{\infty }}
\end{equation}

By the Gagliardo-Nirenberg interpolation inequality (see for instance Aubin
p. 94, CB-DM II p.385) it holds that 
\begin{equation}
||\partial ^{j}u||_{L^{q}}\leq C\{||u||_{W_{2}^{p}}^{\lambda
}||u||_{L^{p}}^{1-\lambda }
\end{equation}
with: 
\begin{equation}
\lambda =\frac{n}{2}\{\frac{j}{n}+\frac{1}{p}-\frac{1}{q}\}
\end{equation}
if 
\begin{equation}
0<\lambda <1.
\end{equation}
Applying the equality 7.27 to the case $j=1,q=n$ gives 
\begin{equation}
0<\lambda =\frac{n}{2p}<1\text{ \ since \ }p>\frac{n}{2}
\end{equation}
Therefore, with this value of $\lambda ,$ it holds that: 
\begin{equation}
||\partial u||_{L^{n}}\leq C||\partial ^{2}u||_{L^{p}}^{\lambda
}||u||_{L^{p}}^{1-\lambda }.
\end{equation}
The same procedure applied to the case $j=1,p=q,\lambda =\frac{1}{2}$ gives
that 
\begin{equation}
||\partial u||_{L^{p}}\leq C||\partial ^{2}u||_{L^{p}}^{\frac{1}{2}%
}||u||_{L^{p}}^{\frac{1}{2}}.
\end{equation}

The interpolation inequality 7.26 with $q=\infty $ and $j=0$ gives, with
again $\lambda =\frac{n}{2p}$: 
\begin{equation}
||u||_{L^{\infty }}\leq C||\partial ^{2}u||_{L^{p}}^{\lambda
}||u||_{L^{p}}^{1-\lambda }.
\end{equation}

For any $\varepsilon >0,$ by elementary calculus, it holds that 
\begin{equation}
||\partial ^{2}u||_{L^{p}}^{\lambda }||u||_{L^{p}}^{1-\lambda }\leq \lambda
\varepsilon ||\partial ^{2}u||_{L^{p}}+\frac{1}{(1-\lambda )\varepsilon
^{\lambda /1-\lambda }}||u||_{L^{p}}
\end{equation}
hence there exists a number $C$ such that 
\begin{equation}
||a_{1}\partial u+a_{0}u||_{L^{p}}\leq
C\{||a_{1}||_{W_{1}^{p}}+||a_{0}||_{L^{p}}\}\{\varepsilon
||u||_{W_{2}^{p}}+C_{\varepsilon }||u||_{L^{p}}\}
\end{equation}
Using this inequality together with 7.20 we see that we can choose $%
\varepsilon >0$ small enough, depending on the bound of $%
||a_{1}||_{W_{1}^{p}}+||a_{0}||_{L^{p}}$ and $C_{a_{2}}$ such that the
inequality 
\begin{equation}
\Vert u\Vert _{W_{2}^{p}}\leq C_{L}\{\Vert Lu\Vert _{L^{p}}+||u||_{L^{p}}\}
\end{equation}
is satisfied. But by the definition of integrals and norms it holds that 
\begin{equation}
||u||_{L^{p}}\leq ||u||_{L^{\infty }}^{(p-1)/p}||u||_{L^{1}}^{1/p}\leq
C||u||_{W_{2}^{p}}^{(p-1)/p}||u||_{L^{1}}^{1/p}
\end{equation}
Therefore if $p>\frac{n}{2}$, for any $\varepsilon >0$ there exists $%
\varepsilon $ and $C_{\varepsilon }$ such that: 
\begin{equation}
||u||_{L^{p}}\leq ||u||_{W_{2}^{p}}^{(p-1)/p}||u||_{L^{1}}^{1/p}\leq
\varepsilon ||u||_{W_{2}^{p}}+C_{\varepsilon }||u||_{L^{1}},
\end{equation}
we can again choose $\varepsilon $ so that the announced inequality holds.
\end{proof}

\begin{proof}
of corollary. This regularity statement can be proved by standard recursive
arguments.
\end{proof}

\begin{theorem}
Under the hypothesis of the theorem 7.2 for $s=0$, and its corollary 7.3 in
the case $s>0,$ it holds that:

1. The operator $L$ maps $W_{2+s}^{p}$ into $W_{s}^{p}$ with finite
dimensional kernel and closed range.

2. If $L$ is injective on $W_{2+s}^{p},$ then there is a number $C_{L}$ such
that for each $u$ in $W_{s+2}^{p}$ the following inequality holds: 
\begin{equation}
\Vert u\Vert _{W_{2+s}^{p}}\leq C_{L}\Vert Lu\Vert _{W_{s}^{p}}\;.
\end{equation}

3.\ If the formal adjoint $^{\ast }L$ of $L$ satisfies the same hypothesis
as $L$ and is injective, then $L$ is surjective from $W_{2+s}^{p}$ onto $%
W_{s}^{p},$ hence an isomorphism if also injective.
\end{theorem}

\begin{proof}
It follows usual lines.
\end{proof}

\subsection{Isomorphism theorem for the Poisson operator.}

The Poisson operator in a riemannian metric $\gamma $ on $(M,e)$ acts on
scalar functions $u$ and reads, with $S_{ij}^{h}$ the difference of the
connections of $e$ and $\gamma :$%
\begin{equation}
\triangle _{\gamma }u-au\equiv \gamma ^{ij}(\partial
_{ij}^{2}u+S_{ij}^{h}\partial _{h}u)-au.
\end{equation}
We suppose that $\gamma \in M_{2}^{p},$ $p>\frac{n}{2},$ then $\gamma \in
C^{0,\alpha }.$ We have denoted by $M_{2}^{p}$ the open subset of symmetric
2-tensors in $W_{2}^{p}$ which are properly riemannian metrics, namely such
that they admit a positive ellipticity constant, $\underset{\cup M_{I}}{Inf}%
\det (\gamma _{ij})>0,$ where $\gamma _{ij}$ are the components of $\gamma $
in a finite number of charts $M_{I}$ covering $M.$ The coefficient $%
a_{2}\equiv \gamma ^{\#},$ the contrevariant tensor associated to $\gamma ,$
is then in $C^{0,\alpha }$ with a positive ellipticity constant. The
coefficient $a_{1}\equiv \gamma ^{ij}S_{ij}^{.}$ is in $W_{1}^{p}$ because $%
\partial \gamma \partial \gamma \in L^{p}$ if $\gamma \in W_{2}^{p},$ $p>%
\frac{n}{2}.$ The hypothesis of the theorem 7.2 are satisfied if $a\in
L^{p}. $ Hence:

\begin{lemma}
If $\gamma \in M_{2}^{p},$ $p>\frac{n}{2}$ , the Poisson operator is a
continuous mapping $W_{2}^{p}\rightarrow L^{p}$ if $a\in L^{p}.$
\end{lemma}

We now prove a uniqueness lemma\footnote{{\footnotesize The conditions p%
\TEXTsymbol{>}n/2 for }$\gamma ${\footnotesize \ and for a are not
necessary, but they are sufficient.}}.

\begin{lemma}
Let $\gamma \in M_{2}^{p},$ $p>\frac{n}{2}$ , $a\in L^{p},$ the Poisson
operator $\triangle _{\gamma }-a$ is injective on $W_{2}^{p}$ if $a\geq 0,$ $%
a\not\equiv 0$ (i.e. $a>0$ on a subset of $M$ of positive measure).
\end{lemma}

\begin{proof}
The proof for $C^{2}$ functions results from the maximum principle if the
coefficients are bounded. In the more general case one proceeds as follows:
on a compact manifold, if $u_{n}$ is a $C^{2}$ function and $\gamma _{n}\in
C^{1}$, the following identity is obtained by a straightforward integration
by parts 
\begin{equation}
\int_{M}u_{n}(\Delta _{\gamma _{n}}u_{n})\mu _{\gamma _{n}}\equiv
-\int_{M}(\gamma _{n}^{ij}\partial _{i}u_{n}\partial _{j}u_{n})\mu _{\gamma
_{n}}
\end{equation}
This identity applied to a $C^{2}$ solution of $\Delta _{\gamma
_{n}}u_{n}-au_{n}=0,$ implies that $\partial _{i}u_{n}\equiv 0$ on $M$, and $%
u_{n}=0$ on the subset $a>0.$ Hence $u_{n}=$constant, and $u_{n}\equiv 0$ .

Suppose now that $\gamma \in M_{2}^{p},$ $p>\frac{n}{2}$, and $u\in
W_{2}^{p}.$ Since $C^{2}$ is dense in $W_{2}^{p}$ we can approximate $\gamma 
$ and $u,$ and $a$ by smooth sequences $\gamma _{n},$ $u_{n}$ converging to $%
\gamma $ and $u,$ respectively in $M_{2}^{p},$ $W_{2}^{p}.$ We already know
by the lemma 7.5 that $\Delta _{\gamma _{n}}u_{n}$ converges in $L^{p}$ to $%
\Delta _{\gamma }u.$ Since the measures $\mu _{\gamma _{n}}$ and $\mu
_{\gamma }$ are equivalent to $\mu _{e}$ and the continuous embedding $%
W_{2}^{p}\times L^{p}\rightarrow L^{1}$ always holds as well as $%
W_{1}^{p}\times W_{1}^{p}\subset L^{1}$\ if $p>\frac{n}{2},$ the integrals
on both sides of 7.30 converge to limits, and it holds that: 
\begin{equation}
\int_{M}u(\Delta _{\gamma }u-au)\mu _{\gamma }\equiv -\int_{M}(\gamma
^{ij}\partial _{i}u\partial _{j}u+au^{2})\mu _{\gamma }
\end{equation}
from which follows $u\equiv 0$ if $\Delta _{\gamma }u-au=0$ and $a\geq
0,a\not\equiv 0.$
\end{proof}

\begin{theorem}
The Poisson operator $\triangle _{\gamma }-a$ on scalar functions in a
metric $\gamma $ on a smooth compact riemannian manifold $(M,e)$, with $%
\gamma \in M_{2}^{p},$ $p>\frac{n}{2},$ $a\in L^{p},$ is an isomorphism from 
$W_{2}^{p}$ onto $L^{p}$ if $a\geq 0,$ $a\not\equiv 0.$

It is an isomorphism $W_{s+2}^{p}\rightarrow W_{s}^{p},$ if in addition $%
a\in W_{s}^{p}.$
\end{theorem}

\begin{proof}
Under the given hypothesis the Poisson operator is selfadjoint relatively to
the metric $\gamma :$ two integrations by part for smooth enough $\gamma ,$ $%
u$ and $v,$ and taking limits, show that under the given hypothesis, as in
the proof of the lemma 7.6, the following identity holds for $u,v\in
W_{2}^{p}:$ 
\begin{equation}
\int_{M}v(\Delta _{\gamma }u-au)\mu _{\gamma }\equiv \int_{M}u(\Delta
_{\gamma }v-av)\mu _{\gamma }.
\end{equation}
The isomorphism theorem 7.4 applies therefore to our Poisson operator.
\end{proof}

\subsection{Generalized maximum principle.}

One knows by the classical maximum principle\textbf{\footnote{{\footnotesize %
Protter and Weinberger 1967, p.64.}}} that if a solution $u\in C^{2}$ of the
inequality with bounded coefficients and $a\geq 0$%
\begin{equation*}
\triangle _{\gamma }u-au\leq 0,\text{ \ [respectively \ }\triangle _{\gamma
}u-au\geq 0]
\end{equation*}
attains a minimum $\lambda \leq 0$ [respectively a maximum $\lambda \geq 0]$
at a point of $M$ then $u\equiv \lambda $ on$\footnote{{\footnotesize Also
if S is a bounded domain of M with smooth boundary a minimum }$\leq 0$ 
{\footnotesize [respectively a maximum }$\geq 0${\footnotesize ]must be
attained on the boundary.}}$ $M$. Therefore $\triangle _{\gamma }u-au\leq 0$
on $M$ implies $u\geq 0$ on $M$ if $a\geq 0$ on $M.$ We will need to obtain
low regularity solutions the following generalization of the maximum
principle.

\begin{lemma}
If $u\in W_{2}^{p}$, $p>\frac{n}{2}$ satisfies the equation 
\begin{equation}
\Delta _{\gamma }u-au=-f
\end{equation}
with $\gamma \in M_{2}^{p},\ a\in L^{p},$ $a\geq k,$ with $k>0$ a number ,
and $\ f\geq 0$ , then $u\geq 0$ on $M.$
\end{lemma}

\begin{proof}
The lemma holds by the classical maximum principle if $u\in C^{2},$ $\gamma
\in M_{3}^{p}$ $a\in C^{0}$. We approximate $\gamma \in M_{2}^{p}$ by $%
\gamma _{n}\in M_{3}^{p},$ $a$ by $a_{n}\in C^{0},$ $a_{n}\geq k$ and $f\in
L^{p}$ by $f_{n}\in C^{0},$ $f_{n}\geq 0.$ We know by the isomorphism
theorem that each equation 
\begin{equation*}
L_{n}\equiv \Delta _{\gamma _{n}}u_{n}-a_{n}u_{n}=-f_{n}
\end{equation*}
has one solution $u_{n}\in W_{4}^{p}\subset C^{2},$ and $u_{n}\geq 0.$ We
know that there exists a number $C_{L}$ depending only on the $W_{2}^{p}$
norm of $\gamma _{n},$ hence of $\gamma ,$ and the $L^{p}$ norm of $a_{n},$
hence of $a,$ such that: 
\begin{equation}
||u_{n}||_{W_{2}^{P}}\leq C_{L}\{||f_{n}||_{L^{p}}+||u_{n}||_{L^{1}}\}.
\end{equation}
We also know that, since $L_{n}\equiv \Delta _{\gamma _{n}}-a_{n}$ is
injective on $W_{2}^{p},$ there exists a number $C_{L_{n}}$ such that: 
\begin{equation}
||u_{n}||_{W_{2}^{P}}\leq C_{L_{n}}||f_{n}||_{L^{p}},
\end{equation}
however the number $C_{L_{n}}$ does not depend only on the $W_{2}^{p}$ and $%
L^{p}$ norms of $\gamma _{n}$ and $a_{n},$ hence is not a priori independent
of $n.$ In order to obtain our positivity result by a continuity argument we
must estimate $C_{L_{n}}$ independently of $n.$ We proceed as follows.

The equality 7.41 shows that a solution $u\in W_{2}^{p},$ $p>\frac{n}{2},$
of 7.43 satisfies the integral equality 
\begin{equation*}
-\int u(\Delta _{\gamma }u-au)\mu _{\gamma }=\int_{M}\gamma ^{ij}\partial
_{i}u\partial _{j}u+au^{2})\mu _{\gamma }=\int_{M}uf\mu _{\gamma }.
\end{equation*}
which implies if $a\geq k>0$ 
\begin{equation}
\int_{M}u^{2}\mu _{\gamma }\leq k^{-1}\int_{M}uf\mu _{\gamma }
\end{equation}
This inequality and the uniform equivalence through a constant $C_{\gamma }$
(depending only on the $W_{2}^{p}$ norm of $\gamma $ and its ellipticity
constant$)$ of the quadratic forms $\gamma $ and $e,$ together with the
H\"{o}lder inequality give: 
\begin{equation}
||u||_{L^{2}}^{2}\leq C_{\gamma ,k}||u||_{L^{q}}||f||_{L^{q^{\prime }}},%
\text{ \ }\frac{1}{q}+\frac{1}{q^{\prime }}=1,\text{ \ }C_{\gamma
,k}=C_{\gamma }k^{-1}
\end{equation}
We have supposed $f\in L^{p}.$

Suppose first that $p\geq 2.$ We have then, since $M$ is compact, $f\in
L^{2} $ with 
\begin{equation}
||f||_{L^{2}}\leq C||f||_{L^{p}}
\end{equation}
We choose then $q^{\prime }=2,$ $q=2$ and we deduce from 7.46 
\begin{equation}
||u||_{L^{1}}\leq C||u||_{L^{2}}\leq C_{\gamma ,k}||f||_{L^{p}}.
\end{equation}
Hence for a solution $u$ of a Poisson equation with coefficients $\gamma \in
M_{2}^{p},$ $a\in L^{p},$ $p>\frac{n}{2},$ $a\geq k>0$ an inequality 
\begin{equation}
||u||_{W_{2}^{P}}\leq C_{L,k}||f||_{L^{p}},
\end{equation}
where $C_{L,k}$ depends only on the $W_{2}^{p}$ and $L^{p}$ norms of $\gamma 
$ and $a,$ and the value of $k.$\newline

Suppose that $p<2$ (inequality compatible with $p>\frac{n}{2}$ only if $%
n<4). $ We then take $q^{\prime }=p,$ but we have $q>2,$ since $%
q^{-1}=1-p^{-1}.$ We use the equality 
\begin{equation}
||u||_{L^{q}}\leq C||u||_{W_{2}^{p}}^{(q-2)/q}||u||_{L^{2}}^{2/q}
\end{equation}
to obtain that 
\begin{equation}
||u||_{L^{2}}^{2-\frac{2}{q}}\leq C||u||_{W_{2}^{p}}^{(q-2)/q}||f||_{L^{p}}
\end{equation}
hence 
\begin{equation}
||u||_{L^{2}}\leq
C||u||_{W_{2}^{p}}^{(q-2)/2(q-1)}||f||_{L^{p}}^{q/2(q-1)}\leq C\varepsilon
||u||_{W_{2}^{p}}+C_{\varepsilon }||f||_{L^{p}}
\end{equation}
from which follows again an inequality of the type 7.50.

We apply a similar inequality to the difference $u_{n}-u_{n-1}$ which
satisfies the equation: 
\begin{equation}
\Delta _{\gamma _{n}}(u_{n}-u_{n-1})-a_{n}(u_{n}-u_{n-1})=F_{n}
\end{equation}
with 
\begin{equation*}
F_{n}\equiv (\Delta _{\gamma _{n}}-\Delta _{\gamma
_{n-1}})u_{n-1}-(f_{n}-f_{n-1})+(a_{n}-a_{n-1})u_{n-1}.
\end{equation*}
The expression of $\Delta _{\gamma }$ as linear in the second and first
derivatives of $u,$ the usual embedding and multiplication properties, and
an inequality of the type 7.50 satisfied by $u_{n-1},$ show that $F_{n}\in
L^{p}$ with an $L^{p}$ bound of the form 
\begin{equation}
||F_{n}||_{L^{P}}\leq C_{\gamma ,f,k}\{||\gamma _{n}-\gamma
_{n-1}||_{W_{\sigma
}^{p}}+||f_{n}-f_{n-1}||_{L^{p}}+||a_{n}-a_{n-1}||_{L^{p}}\}.
\end{equation}
where $C_{\gamma ,f,k}$ denotes any positive number depending only on $k,$
the $W_{2}^{p}$ norm of $\gamma ,$ and the $L^{p}$ norms of $a$ and $f.$ The
inequality 7.50 applied to 7.54 gives therefore an inequality of the form 
\begin{equation*}
||u_{n}-u_{n-1}||_{W_{2}^{p}}\leq C_{\gamma ,f,k}\{||\gamma _{n}-\gamma
_{n-1}||_{W_{2}^{p}}+||f_{n}-f_{n-1}||_{L^{p}}+||a_{n}-a_{n-1}||_{L^{p}}\}.
\end{equation*}
We deduce from this inequality that the sequence $u_{n}$ converges in $%
W_{2}^{p},$ hence in $C^{0},$ to a function $u\in W_{2}^{p}$ when $f_{n}$
converges to $f$ and $\gamma _{n}$ to $\gamma .$ This function $u$ is the
unique solution of the Poisson equation, and since $u_{n}\geq 0,$ the same
is true of $u\footnote{{\footnotesize Recall that it is to obtain a family
of smooth functions u}$_{n}${\footnotesize \ \ approaching u that we had to
introduce the positive k. Generalized maximum principles can probably be
obtained by other methods.}}.$
\end{proof}

\section{Appendix B. Equation $\Delta _{\protect\gamma }\protect\varphi =f(x,%
\protect\varphi ).$}

We consider equations of the form, with $\varphi $ a scalar function 
\begin{equation}
\Delta _{\gamma }\varphi =f(.,\varphi ).
\end{equation}
where $f:(x,y)\mapsto f(x,y)$ is a function on $M\times I,$ with $I$ an
interval of $R,$ smooth in $y.$ To be specific, and in view of application
to the Lichnerowicz equation, we suppose that $f$ is a finite sum: 
\begin{equation}
f(x,y)\equiv \sum_{I=1,...N}a_{I}(x)y^{P_{I}},
\end{equation}
where the exponents $P_{I}$ are given real numbers. Hence $f$ is a smooth
function of $y$ if $y\geq \ell >0.$

It is to solve this semi-linear equation that we have supposed $p>\frac{n}{2}%
.$ The space $W_{2}^{p}$ is then an algebra, $\varphi ^{P}$ is in $%
W_{2}^{p}\subset C^{0}$ for any $P$ if it is so of $\varphi $ and if $%
\varphi $ is positive.

To solve 8.1 we apply a method\footnote{{\footnotesize The solution of
equations of the indicated type on compact manifolds was first obtained, in
Holder spaces, by using the Leray-Shauder degree theory, together with
bounds by sub and super solutions in CB and Leray 1972. It was applied to
the constraints by CB 1972 (unscaled sources), O'Murchada and York 1974
(scaled sources), extended to H}$_{s}\ ${\footnotesize spaces in CB 1975.
The iteration method (Kazdan and Warner 1985) was applied to the constraints
by Isenberg 1987. \ In the following we weaken the regularity hypothesis
previously made.}} of successive iterations obtained by resolution of linear
problems, whose bounds and convergence are obtained through the use of sub
and super solutions.

\begin{definition}
A $W_{2}^{p}$ function $\varphi _{-}$ is called a \textbf{subsolution} of $%
\Delta _{\gamma }\varphi =f(.,\varphi )$ if it is such that on\footnote{%
{\footnotesize Inequality satisfied for functions in L}$^{p}${\footnotesize %
, i.e. almost everywhere}} $M$%
\begin{equation}
\Delta _{\gamma }\varphi _{-}\geq f(.,\varphi _{-}).
\end{equation}
\smallskip An $W_{2}^{p}$ function $\varphi _{+}$ is called a supersolution
if 
\begin{equation}
\Delta _{\gamma }\varphi _{+}\leq f(x,\varphi _{+}).
\end{equation}
\end{definition}

\begin{theorem}
(existence) The equation $\Delta _{\gamma }\varphi =f(.,\varphi )$ on the
compact manifold (M,$\gamma )$ admits a solution $\varphi \in W_{2}^{p},$ $p>%
\frac{n}{2},$ if:

a. $\gamma \in M_{2}^{p},$ $p>\frac{n}{2},$ a$_{I}\in L^{p},$ I=1,...,N.

b. The equation admits a subsolution $\varphi _{-}$ and a supersolution $%
\varphi _{+},$ both in $W_{2}^{p},$ $0<\ell \leq \varphi _{-}\leq \varphi
_{+}\leq m$. The solution is such that $\varphi _{-}\leq \varphi \leq
\varphi _{+}.$
\end{theorem}

\begin{proof}
The proof follows the same lines as the standard proof, but uses the \
generalized maximum principle (lemma 7.8). We define successive iterates by
solution of the elliptic equation, linear in $\varphi _{n}$ when $\varphi
_{n-1}$ is known, 
\begin{equation}
\Delta _{\gamma }\varphi _{n}-a\varphi _{n}=f(.,\varphi _{n-1})-a\varphi
_{n-1},
\end{equation}
with $a\in L^{p},$ $a\geq k,$ $k$ a strictly positive number to be
determined later, introduced now to allow the application of the theorem 7.7.

We take $\varphi _{0}=\varphi _{-}$ to start the iteration, i.e. we define $%
\varphi _{1}$. by solving the equation 
\begin{equation}
\Delta _{\gamma }\varphi _{1}-a\varphi _{1}=f(.,\varphi _{-})-a\varphi _{-},
\end{equation}
due to the hypothesis made on $\gamma ,a,f$ and $\varphi _{-}$, this
solution $\varphi _{1}$ is in $W_{2}^{p}.$ Moreover we have by the
definitions of $\varphi _{1}$ and $\varphi _{-}$ 
\begin{equation}
\Delta _{\gamma }(\varphi _{1}-\varphi _{-})-a(\varphi _{1}-\varphi
_{-})\leq 0
\end{equation}
\smallskip The conclusion $\varphi _{1}-\varphi _{-}\geq 0$ on $M$ follows
from the generalized maximum principle.

We now use the definition of $\varphi _{+}$ to obtain: 
\begin{equation}
\Delta _{\gamma }(\varphi _{+}-\varphi _{1})-a(\varphi _{+}-\varphi
_{1})\leq f(x,\varphi _{+})-f(x,\varphi _{-})-a(\varphi _{+}-\varphi _{-})
\end{equation}
\smallskip which gives using the mean value theorem (CB-DM I p.78) 
\begin{equation}
\Delta _{\gamma }(\varphi _{+}-\varphi _{1})-a(\varphi _{+}-\varphi
_{1})\leq (\varphi _{+}-\varphi _{-})\int_{0}^{1}\{f_{\varphi }^{\prime
}(x,\varphi _{-}+t(\varphi _{+}-\varphi _{-}))-a\}dt
\end{equation}
It is sufficient to choose the function $a$ such that 
\begin{equation}
a=Sup_{l\leq y\leq m}|f_{y}^{\prime }(.,y)|+k,\text{ \ \ }k>0
\end{equation}
to have on $M$ that \ $a\geq k>0$ and 
\begin{equation}
\Delta _{\gamma }(\varphi _{+}-\varphi _{1})-a(\varphi _{+}-\varphi
_{1})=F\leq 0,\text{ }
\end{equation}
since on $M$ it holds that $0<\ell \leq \varphi _{-}+t(\varphi _{+}-\varphi
_{-})\leq m.$ For example if the function $f$ is given by 8.2, then: 
\begin{equation}
|f_{y}^{\prime }(.,y)|\leq \sum_{I=1,...N}|a_{I}P_{I}|y^{P_{I}-1}\leq
\sum_{I_{+}}|a_{I_{+}}P_{I_{+}}|m^{P_{I_{+}}-1}+%
\sum_{I_{-}}|a_{I_{-}}P_{I_{-}}|\ell ^{P_{I_{-}}-1}+|a_{I_{0}}|
\end{equation}
with $P_{I_{+}}>1,$ $P_{I_{-}}<1,$ $P_{I_{0}}=1.$ Under the hypothesis made
on $a_{I}$ the maximum principle holds, hence $\varphi _{1}\leq \varphi _{+}$
on $M.$

Suppose that there exists $\varphi _{m}\in W_{2}^{p},$ $m=1,...,n-1,$ such
that on $M$ 
\begin{equation}
\varphi _{-}\leq \varphi _{m}\leq \varphi _{+}
\end{equation}
then $\varphi _{n\text{ }}$ exists on $M,$ solution of 8.5, also in $%
W_{2}^{p}$. Suppose moreover that 
\begin{equation}
\varphi _{n-1}-\varphi _{n-2}\geq 0
\end{equation}
Then the equations satisfied by $\varphi _{n}$ and $\varphi _{n-1}$ imply 
\begin{equation}
\Delta _{\gamma }(\varphi _{n}-\varphi _{n-1})-a(\varphi _{n}-\varphi
_{n-1})=(b_{n}-a)(\varphi _{n-1}-\varphi _{n-2})
\end{equation}
where $b_{n}$ is the function in $L^{p}$ 
\begin{equation}
b_{n}\equiv \int_{0}^{1}f_{\varphi }^{\prime }(.,\varphi _{n-2}+t(\varphi
_{n-1}-\varphi _{n-2})dt.
\end{equation}
It results from the definition of $a$ that: 
\begin{equation}
b_{n}-a\leq -k<0
\end{equation}
this inequality together with $\varphi _{n-1}\geq \varphi _{n-2}$\ insures
that the right hand side of (8.15) is negative, hence $\varphi _{-}\leq
\varphi _{n-1}\leq \varphi _{n}$. To show that the sequence is bounded above
by $\varphi _{+}$ we write again an inequality of the form 
\begin{equation}
\Delta _{\gamma }(\varphi _{+}-\varphi _{n})-a(\varphi _{+}-\varphi
_{n})\leq f(\varphi _{+})-f(\varphi _{n-1})-a(\varphi _{+}-\varphi _{n-1})
\end{equation}
\smallskip 
\begin{equation*}
\leq (b_{n}-a)(\varphi _{+}-\varphi _{n-1})\leq 0,
\end{equation*}
\smallskip from which the conclusion follows. We have proved the existence
of the sequence $\varphi _{n}\in W_{2}^{p},$ with $0\leq \varphi _{-}\leq $ $%
\varphi _{n}\leq \varphi _{+}$ hence \TEXTsymbol{\vert}\TEXTsymbol{\vert}$%
\varphi _{n}||_{L^{p}}$ uniformly bounded by \TEXTsymbol{\vert}\TEXTsymbol{%
\vert}$\varphi _{+}||_{L^{p}}$. Due to the defining equation 8.5 $||\varphi
_{n}||_{W_{2}^{p}}$ is also uniformly bounded. Since functions in $W_{2}^{p}$
are equicontinuous we can extract (Ascoli - Arzela theorem) from the
sequence $\varphi _{n}$ a subsequence, $\tilde{\varphi}_{n},$ which
converges in $C^{0}$ norm, to a function $\varphi \in C^{0}.$ On the other
hand the sequence $\varphi _{n}$ of continuous and positive functions is
increasing and bounded above by $\varphi _{+},$ hence it is pointwise
convergent to a funtion $\varphi ,$ which coincides with the previously
obtained $\varphi $. The whole sequence (not only a subsequence) is
therefore convergent.

We show that the limit is in fact in $W_{2\text{ }}^{p}$ by using the
elliptic estimate applied to the equation 8.15 
\begin{equation}
\parallel \varphi _{n}-\varphi _{n-1}\parallel _{W_{2}^{p}}\leq C_{\gamma
,a}||(b_{n}-a)(\varphi _{n-1}-\varphi _{n-2})\parallel _{L^{p}}
\end{equation}
which implies: 
\begin{equation}
\parallel \varphi _{n}-\varphi _{n-1}\parallel _{W_{2}^{p}}\leq C_{\gamma
,a}||(b_{n}-a)||_{L^{p}}||\varphi _{n-1}-\varphi _{n-2})\parallel _{C^{0}},
\end{equation}
The sequence $\varphi _{n}$ converges in $W_{2}^{p}$ (to $\varphi $ by
uniqueness of limits) since it converges in $C^{0},$ and the $b_{n}-a$ are
uniformly bounded in $L^{p}$ norm. The limit satisfies the equation in $%
L^{p} $ sense.

The solution we have constructed depends on the choice of the initial $%
\varphi _{0}$. We have taken $\varphi _{-}$ as an initial $\varphi _{0}$. We
could have taken $\varphi _{+}$ and construct a decreasing and bounded below
sequence which converges to a limit $\Psi $. We cannot in the procedure used
start from an arbitrary $\varphi _{0}$, with $\varphi _{-}\leq \varphi
_{0}\leq \varphi _{+}$, because we will not know if $\varphi _{1}$ satisfies
the same inequalities. A uniqueness theorem is easy to prove in the case
where $y\mapsto f(.,y)$ is monotonically increasing. A more general
uniqueness theorem (4.4) holds for the Lichnerowicz equation.
\end{proof}

\textbf{Aknowledgement. }\bigskip I thank the referee for useful
criticisisms.

\textbf{Bibliography.}

Aubin A. Non linear analysis on manifolds. Monge Ampere equations,
Springer-Verlag

Bartnik R. 1988 Remarks on cosmological spacetimes and constant mean
curvature surfaces. Comm Math. Phys. 117, 615-624.

Choquet (Foures) - Bruhat Y. (1948) Sur l'int\'{e}gration du probl\`{e}me
des conditions initiales en m\'{e}canique relativiste C.R. Acad. Sci. 226,
1071-1073.

Choquet (Foures) - Bruhat Y. (1956) Sur l'int\'{e}gration des \'{e}quations
de la relativit\'{e} g\'{e}n\'{e}rale, J. Rat. Mech. Anal. 5, 951-966.

Choquet (Foures) - Bruhat Y. (1957) Sur le probl\`{e}me des conditions
initiales C.R. Acad. Sci. 245, see also Bruhat Y.\ (1962) The Cauchy
problem, in ''Gravitation an introduction to current research'', L. Witten
ed, Wiley.

Choquet - Bruhat Y.\ (1970) A new elliptic system and global solutions for
the constraint equations in General Relativity. Comm. Math. Phys. 21,
211-218.

Choquet-Bruhat Y. (1972) Solutions globales du probl\`{e}me des contraintes
sur une vari\'{e}t\'{e} compacte. C.R. Acad. Sci. 274, 682-684, Global
solutions of the problem of constraints on closed manifolds, Acta Ist. alta
mat. XII, 317-325.

Choquet-Bruhat Y. (1975) Solution of the constraints in Sobolev space $%
H_{4}, $ volume in honor of Ivanenko.

Choquet-Bruhat Y. (1996) In Gravity particles and fields, Sardanishvily ed.
World Scientific 19-28.

Choquet-Bruhat Y. , DeWitt-Morette C. (1982) Analysis, Manifolds and Physics
I North Holland.

Choquet-Bruhat Y., Isenberg J. and Moncrief V. (1992) Solutions of
constraints for Einstein equations C.R. Acad. Sci. 315, 349-355.

Choquet-Bruhat Y. , Isenberg J. , York J.W. (2000) Einstein constraints on
asymptotically euclidean manifolds Phys Rev D (3) 61 n$%
{{}^\circ}%
8.$

Choquet-Bruhat Y. and Leray J. (1972) Sur le probl\`{e}me de Dirichlet quasi
lin\'{e}aire d'ordre 2. C.R. Acad. Sci. 274, 81-85.

Choquet-Bruhat Y. and Moncrief V. Existence theorems for solutions of
Einstein's equations with 1 parameter isometry group. Proc. symp. pure Math.
59, 67-80.

Choquet-Bruhat Y. and York J.W. (1980) Cauchy problem, in General Relativity
and Gravitation, Held ed. Plenum

Fisher A.\ E. and Moncrief V. (1994) Reducing Einstein equations to an
unconstrained Hamiltonian system on the cotangent bundle of Teichm\"{u}ller
space, in Physics on manifolds, M.\ Flato, R.\ Kerner, A.\ Lichnerowicz ed.
Kluwer.

Isenberg J. (1987) Parametrization of the space of solutions of Einstein
equarions, Phys. Rev. Let. 59, 2389-2392.

Isenberg J. (1995) Constant mean curvature solutions of Einstein constraint
equations on closed manifolds, Class. Quant. Grav. 12, 2249-2279.

Isenberg J. and Moncrief V.\ (1996) A set of nonconstant mean curvature
solutions of Einstein constraint equations on closed manifolds, Class.
Quant. Grav. 13, 1819-1847.

Lichnerowicz A.\ (1944) L'int\'{e}gration des \'{e}quations relativistes et
le probl\`{e}me des n corps J.\ Math. pures et app. 23, 37-63.

Lichnerowicz A. (1963) Spineurs harmoniques, C.R. Acad. Sci.257, 7-9.

Moncrief V. (1975) Spacetime symmetries and linearization stability of the
Einstein equations. J.\ Math. Phys 16 493-498.

Moncrief V. (1986) Reduction of Einstein spacetimes with U(1) isometry group
Ann of Phys 167 118-142.

O'Murchada N. and York J.W. (1973) Existence and uniqueness of solutions of
the Hamiltonian constraint of General Relativity on compact manifolds J.
Math. Phys. 14, 1551-1557.

O'Murchada N. and York J.W. (1974) Initial value problem of general
relativity, Phys. Rev. D 10 n$%
{{}^\circ}%
2$, 428-436.

O'Murchada N. and York J.W. (1974) Initial value problem of general
relativity II, stability of solutions, Phys. Rev. D 10 n$%
{{}^\circ}%
2$, 437-446.

Pfeister M. and York J.W. (2002) Phys Rev D

York J.W. (1972) Role of conformal 3 - geometry in the dynamics of
gravitation; Phys Rev Lett 28, 1082

York J.W. (1999) Conformal thin sandwich initial data for the initial value
problem of general relativity. Phys Rev Lett. 82, 1350-1353.

\end{document}